\documentclass[twocolumn,aps,showpacs,prb,tightenlines,amsmath,amssymb]{revtex4}
\usepackage{graphicx}
\usepackage{amssymb}
\usepackage{slashed}
\usepackage{dcolumn}
\usepackage{amsmath}
\usepackage{bm}
\usepackage{colordvi}
\usepackage{mathrsfs}
\makeatletter

\newcommand{\Rmnum}[1]{\expandafter\@slowromancap\romannumeral #1@}
\makeatother

\begin{document}

\title{Gauge-invariant microscopic kinetic theory of superconductivity:
  application to electromagnetic response of Nambu-Goldstone and Higgs modes}   

\author{F. Yang}
\affiliation{Hefei National Laboratory for Physical Sciences at
Microscale, Department of Physics, and CAS Key Laboratory of Strongly-Coupled
Quantum Matter Physics, University of Science and Technology of China, Hefei,
Anhui, 230026, China}

\author{M. W. Wu}
\thanks{Author to whom correspondence should be addressed}
\email{mwwu@ustc.edu.cn.}

\affiliation{Hefei National Laboratory for Physical Sciences at
Microscale, Department of Physics, and CAS Key Laboratory of Strongly-Coupled
Quantum Matter Physics, University of Science and Technology of China, Hefei,
Anhui, 230026, China}

\date{\today}

\begin{abstract} 

We show that the gauge-invariant kinetic equation of superconductivity provides an efficient approach to study the
electromagnetic response of the gapless Nambu-Goldstone and gapful Higgs modes on an equal footing.
We prove that the Fock energy in the kinetic equation is equivalent to the generalized Ward's identity. Hence, 
the gauge invariance directly leads to the charge conservation. Both linear and second-order responses are investigated.   
The linear response of the Higgs mode vanishes in the long-wave limit. Whereas the linear response of 
the Nambu-Goldstone mode interacts with the long-range Coulomb interaction, causing the original gapless spectrum
lifted up to the plasma frequency as a result of the Anderson-Higgs mechanism, in consistency with the previous
works.  The second-order response exhibits interesting physics. On one hand, a finite second-order response of the Higgs
mode is obtained in the long-wave limit. We reveal that this response, which has been experimentally observed, is
attributed solely to the drive effect rather than the widely considered Anderson-pump effect. On the other hand, the
second-order response of the Nambu-Goldstone mode,   free from the influence of the long-range Coulomb interaction and
hence the Anderson-Higgs mechanism, is predicted.  We find that both Anderson-pump and drive effects play
important role in this response. A tentative scheme to detect this second-order response is proposed.   

\end{abstract}


\maketitle

\section{Introduction}

The collective excitation in the superconducting states has been the focus of study in
the field of superconductivity for the past few 
decades. Two types of collective modes emerge with the 
generation of the superconducting order parameter $\Delta$: the gapless phase
mode\cite{gi0,AK,Gm1,Gm2,Ba0,pm0,pi1,pm1,pm2,pi4,gi1,Ba9,Ba10,pm5} and
gapful amplitude
mode,\cite{pm5,Am0,Am5,Am6,Am12,Am13}
which correspond to the fluctuation of phase and amplitude of the order
parameter, respectively. Specifically, through the generalized Ward's
identity, Nambu first revealed the existence of a collective gapless excitation
in the superconducting states.\cite{gi0} It is understood later that
this gapless excitation is described as a collective phase mode of the order
parameter\cite{AK} and corresponds to the gapless Goldstone bosons in the field
theory by the spontaneous breaking of the continuous $U(1)$ symmetry.\cite{Gm1,Gm2} 
After that, the phase mode is further proved by obtaining the effective
Lagrangian of the order parameter via the path integral
method,\cite{pi1,pi4} and is now referred to as the Nambu-Goldstone (NG) 
mode.\cite{pm1,pm2,pi4,gi1,Ba9,Ba10,pm5} The counterpart of the phase mode is the
amplitude
mode,\cite{pm5,Am0,Am6,Am12,Am13}
which is referred to as the Higgs mode due to the similarity of the
Higgs bosons in the field theory.\cite{Higgs1,Higgs2,Higgs3} Particularly, a
gapful energy spectrum $\omega_{\rm H}=2|\Delta|$ of the Higgs mode in
superconductors is predicted in the long-wave limit.\cite{pm5,Am0,Am6,Am13}

Since the elucidation of the existence of the collective modes in superconductors, 
a great deal of theoretical efforts have been devoted to their electromagnetic response. 
Nevertheless, the theoretical studies of the electromagnetic responses of the NG
mode and Higgs mode in the literature so far are separated  by either fixing the
amplitude or overlooking the
phase of the order
parameter.   
Moreover, due to the spontaneous breaking of the $U(1)$ symmetry by the
generation of the order parameter,  it is
established\cite{gi0,Ba0,gi1,Ba9,Ba10} that the gauge transformation in the
superconducting states contains the superconducting phase $\theta$ of the order
parameter, in addition to the standard electromagnetic potential $A_{\mu}=(\phi,{\bf
  A})$. Nambu pointed out\cite{gi0,gi1,Ba0} via the generalized Ward's identity
that the gauge invariance in the superconducting states is equivalent to the
charge conservation. Since the charge conservation is directly related to the
electromagnetic properties, the gauge invariance is necessary for the
physical description. Nevertheless, a complete gauge-invariant theory for the electromagnetic response of both
collective modes is still in progress.       

Specifically, with the fixed amplitude of the order parameter, via the Gorkov's equation,\cite{G1} it is first 
revealed by Ambegaokar and Kadanoff\cite{AK} that the NG mode
responds to the electromagnetic field in the linear
regime. Nevertheless, this linear response of the NG mode   
interacts with the long-range Coulomb interaction,\cite{AK} causing the original
gapless energy spectrum lifted up to the high-energy plasma frequency as a
result of Anderson-Higgs mechanism.\cite{AHM}  However, for the
gauge-invariant approach in Ambegaokar and Kadanoff's work,\cite{AK} in order to obtain the NG
mode, an additional condition of the charge conservation is 
required.\cite{AKs} This seems superfluous since as mentioned above, the
presence of the gauge invariance directly implies the charge
conservation.\cite{gi0,gi1,Ba0} After that, the Anderson-Higgs mechanism of the
NG mode in the linear response is further discussed within the diagrammatic
formalism.\cite{pm0,pm5,gi1,Ba0,Am0,gi0} However, due to the difficulty in treating
the nonlinear effect in the diagrammatic formalism or
the Gorkov's equation, the nonlinear response of the NG mode is absent
in the literature.   

The electromagnetic response of the Higgs mode has recently been focused  
in the second-order regime. This is inspired by the recent
experiments,\cite{NL7,NL8,NL9,NL10,NL11} from which it is
realized that the intense THz field can excite the oscillation of the superfluid 
density in the second-order response. This oscillation so far is attributed
to the excitation of the Higgs mode based on the observed resonance when
the optical frequency is tuned at the superconducting gap.\cite{NL8,NL9,NL10} 
Theoretical description for this response
has been based on the Bloch\cite{Am1,Am2,Am7,Am9,Am11,Am14,NL7,NL8,NL9,NL10,NL11} or
Liouville\cite{Am3,Am4,Am8,Am10} equation derived in the Anderson
pseudospin picture.\cite{As} The second-order term $A^2$ naturally emerges in these
descriptions,\cite{Am1,Am2,Am3,Am4,Am7,Am8,Am9,Am10,Am11,Am14,NL7,NL8,NL9,NL10,NL11} causing the 
pump of the quasiparticle correlation (pump effect) and hence the
fluctuation of the order parameter $\Delta$. Then, it is claimed that
the Higgs mode is excited. Recently, this description is
challenged.\cite{symmetry,GOBE1,GOBE2,GOBE3,GOBE4} Firstly, the symmetry 
analysis\cite{symmetry} from the Anderson pseudospin 
picture implies that with the particle-hole symmetry,
the excited fluctuation of the order parameter by the pump
effect is the oscillation of its phase. This suggests that the pump effect excites the
NG mode rather than the Higgs mode. Secondly, the Bloch\cite{Am1,Am2,Am7,Am9,Am11,Am14,NL7,NL8,NL9,NL10,NL11} or
Liouville\cite{Am3,Am4,Am8,Am10} equation fails in
the linear response to describe the optical conductivity since no drive effect
(i.e., linear term) is included.\cite{GOBE3,GOBE4} Thus, these descriptions 
are insufficient to elucidate the complete
physics. Most importantly, with the vector potential alone, the gauge
invariance is unsatisfied in the Bloch or Liouville equation in the
literature.\cite{GOBE1,GOBE2,GOBE3,GOBE4}

Very recently, we extend the nonequilibrium $\tau_0$-Green-function approach ($\tau_i$ are the Pauli matrices in the
particle-hole space), which has been very successful in studying the dynamics of the semiconductor 
optics\cite{GQ2} and spintronics,\cite{GQ3}  into the dynamics of superconductivity. The equal-time scheme in this
approach, corresponding to the instantaneous optical transition\cite{GQ2} in optics and the nonretarded spin
precession\cite{GQ3} in spintronics, can naturally be applied into the conventional $s$-wave superconducting states  
thanks to the BCS equal-time pairing.\cite{BCS} To retain the gauge invariance, the gauge-invariant $\tau_0$-Green
function is constructed through the Wilson line\cite{Wilson}. Then, a gauge-invariant kinetic equation (GIKE) is developed
for the electromagnetic response of the superconductivity.  As a result of the gauge invariance, both the
Anderson-pump and drive effects mentioned above are kept. By following the previous
approaches,\cite{Am1,Am2,Am3,Am4,Am5,Am6,Am7,Am8,Am9,Am10,Am11,Am12,Am13,Am14,NL7,NL8,NL9,NL10,NL11} i.e., overlooking   
the NG mode (lifted up to the plasma frequency by the long-range Coulomb interaction), it is shown by the
GIKE\cite{GOBE1,GOBE2,GOBE3,GOBE4} that instead of the well-studied Anderson-pump effect in the
literature,\cite{Am1,Am2,Am7,Am9,Am11,Am14,NL7,NL8,NL9,NL10,NL11,Am3,Am4,Am8,Am10} 
the second order of the drive effect dominates the second-order response of the Higgs mode.
Moreover, in a latest paper,\cite{GOBE4} we further show that both superfluid and normal-fluid dynamics are involved
in the GIKE, beyond the Boltzmann equation of superconductors in the literature\cite{Ba3,Bol,Ba5} which
only includes the quasiparticle excitations.   
Particularly, the equal-time scheme in the GIKE makes it very easy to handle
the temporal evolution and microscopic scattering in the superconducting states,  in contrast to the conventional
Eilenberger transport equation in superconductors which is derived from $\tau_3$-Green function and restricted by the
normalization condition.\cite{Eilen,Ba7,Ba8,Ba20}  Consequently, in addition to the well-known clean-limit results such
as the Ginzburg-Landau equation near $T_c$ and the Meissner supercurrent in the magnetic response, from the GIKE, rich
physics by the microscopic scattering has been revealed.\cite{GOBE4} Specifically,  we find there exists a friction
between the normal-fluid and superfluid currents and  due to this friction, part of the superfluid becomes viscous.
Therefore, a three-fluid model with the normal fluid and nonviscous and viscous superfluids is proposed.

In this work, we show that the GIKE developed before\cite{GOBE1,GOBE2,GOBE3,GOBE4} also provides an efficient
approach to study the electromagnetic response of the collective modes in the superconducting states.  
We first demonstrate that the generalized Ward's identity by Nambu\cite{gi0,Ba0} is equivalent to the Fock
energy in the GIKE.   With the complete Fock term, the gauge invariance in the GIKE
directly leads to the charge conservation, in contrast to the previous Ambegaokar and
Kadanoff's approach\cite{AK} where an additional condition of the charge conservation is required to
obtain the NG mode.  In addition to the Fock term in our previous GIKE,\cite{GOBE4} the Hartree one (i.e., the vacuum
polarization) is also added in the present work.  Then, both linear and second-order responses are investigated.
Differing from the previous studies in the literature with either the fixed amplitude\cite{AK,pi1,pm1,pm2,pi4,Ba9}
or overlooked
phase\cite{Am1,Am2,Am3,Am4,Am5,Am6,Am7,Am8,Am9,Am10,Am11,Am12,Am13,Am14,NL7,NL8,NL9,NL10,NL11,GOBE1,GOBE2,GOBE3,GOBE4}
of the order  parameter, in the present work, the gapless NG and gapful Higgs modes are calculated on an equal
footing. 
Consequently, the contributions from the phase and amplitude modes to the fluctuation of the order parameter,  which
are ambiguous in the Anderson pseudospin picture as mentioned above, can be directly distinguished in our
GIKE approach.

Specifically, the linear response of the NG mode from our GIKE agrees with the previous results in the
literature.\cite{AK,Ba0,pm0,Am0,Ba9,Ba10,pm5}  The linear response of the NG mode interacts with the long-range Coulomb
interaction, causing the original gapless energy spectrum inside the superconducting gap effectively lifted up to the
high-energy plasma frequency far above the gap as a result of Anderson-Higgs mechanism.\cite{AHM} Consequently,
no effective linear response of the NG mode occurs. The origin of the plasma frequency is addressed. The second-order
response of the NG mode to the electromagnetic field,  which is hard to deal with in the previous 
approaches in the literature,\cite{AK,Ba0,Ba9,Ba10,pm0,pm5,Am0,pi1,pi4} exhibits interesting physics in contrast to the
linear one. Specifically, in the second-order regime,  we find that the NG mode also responds to the electromagnetic field.
Both the Anderson-pump effect and the second order of the drive effect play important role.  
Particularly, in striking contrast to the linear response above,  it is very interesting to find that the
second-order response of the NG mode decouples with the long-range Coulomb interaction, and hence maintains the
original gapless energy spectrum inside the superconducting gap, free from the influence of  the Anderson-Higgs
mechanism.\cite{AHM}  The origin of this decoupling is revealed. A
tentative scheme to detect the second-order response of the NG mode is also proposed. 

As for the Higgs mode, we find that the Higgs mode also responds to the electromagnetic field in the linear    
regime but this response vanishes in the long-wave limit. A finite response of the Higgs mode in the long-wave limit
is obtained in the second-order regime. By further comparing the Anderson-pump and drive effects, we show that the widely
considered pump effect in the
literature\cite{Am1,Am2,Am7,Am9,Am11,Am14,Am3,Am4,Am8,Am10,NL7,NL8,NL9,NL10,NL11,GOBE1,GOBE2,GOBE3,GOBE4} makes no
contribution at all.  Only the second order of the drive effect contributes to the second-order response of the Higgs
mode, and exhibits a resonance at $2\omega=2\Delta$, in  
consistency with the experimental findings.\cite{NL8,NL9,NL10}
Consequently,  the experimentally observed second-order response of the Higgs mode is attributed
solely to the drive effect rather than the pump effect widely speculated in the
literature.\cite{Am1,Am2,Am7,Am9,Am11,Am14,Am3,Am4,Am8,Am10,NL7,NL8,NL9,NL10,NL11,GOBE1,GOBE2,GOBE3,GOBE4} The pump
effect only contributes to the second-order response of the NG mode as mentioned above.  

This paper is organized as follows. We first present the Hamiltonian and introduce the GIKE of superconductivity in
Sec.~II~A and~B, respectively. Then, we show in Sec.~II~C that the generalized Ward's identity by Nambu is equivalent to 
the Fock energy in the GIKE. The demonstration of the charge conservation from the GIKE is addressed in
Sec.~II~C. We perform the analytical analysis for the electromagnetic response of the Higgs and NG modes in the linear
and second-order regimes in Sec.~III. We summarize in Sec. IV.

\section{MODEL}
\label{model}

In this section, we first present the Hamiltonian of the conventional
superconducting states and the corresponding gauge structure revealed by
Nambu.\cite{gi0,gi1} Then, we introduce the GIKE of the superconductivity and prove the charge conservation from the GIKE.    

\subsection{Hamiltonian}

In the presence of the electromagnetic field, the Bogoliubov-de Gennes (BdG)
Hamiltonian of the conventional superconducting states is written as   
\begin{equation}
\label{BdG}
H={\int}\frac{d{\bf r}}{2}\psi^{\dagger}(x)\{[\xi_{{{\bf p}}-e{\bf
    A}(x)\tau_3}+e\phi(x)]\tau_3+{\hat \Sigma}(x)\}\psi(x), 
\end{equation}
with the Fock energy in the BCS pairing scheme:
\begin{equation}
{\hat \Sigma}(x)=\left(\begin{array}{cc}
\mu_0+\mu_F(x) & |\Delta(x)|e^{i\theta(x)}\\
|\Delta(x)|e^{-i\theta(x)} & \mu_0-\mu_F(x)\\
\end{array}\right).
\end{equation}
Here, $\psi(x)=[\psi_{\uparrow}(x),\psi^{\dagger}_{\downarrow}(x)]^T$ is
the field operator in the Nambu space; 
$\xi_{{\bf p}}={{\bf p}^2}/({2m})-\mu$ with $m$ and $\mu$  
being the effective mass and chemical potential; ${{\bf p}}=-i\hbar{\bm {\nabla}}$; $\mu_F$ stands for the Fock field;
$|\Delta|$ and $\theta$ represent the amplitude and phase of the order parameter, respectively.  

Due to the spontaneous breaking of the $U(1)$ symmetry by the generation
of the superconducting order parameter $\Delta$, the gauge transformation in
superconductors reads:\cite{gi0,Ba0,Ba9,Ba10,gi1}  
\begin{eqnarray}
eA_{\mu}&\rightarrow&eA_{\mu}-\partial_{\mu}\chi(x), \label{gaugestructure1}\\
\label{gaugestructure2}
\theta(x)&\rightarrow&\theta(x)+2\chi(x),
\end{eqnarray}
with the four vector $\partial_{\mu}=(\partial_t,-{\bm \nabla})$.

\subsection{Gauge-invariant microscopic kinetic theory}

By adding the Hartree term (i.e., the vacuum polarization) into our previous GIKE,\cite{GOBE4} the new GIKE reads:
\begin{eqnarray}
&&\partial_T\rho^c_{\bf
  k}\!+\!i\Big[(\xi_k\!+\!e\phi\!+\!\mu_H)\tau_3\!+\!{\hat
  \Sigma}_F(R),\rho^c_{\bf k}\Big]\!+\!i\Big[\frac{e^2A^2}{2m}\tau_3,\rho^c_{\bf k}\Big]\nonumber\\
&&\mbox{}\!+\!\frac{1}{2}\left\{e{\bf E}\tau_3\!-\!({\bm \nabla}_{\bf R}\!-\!2ie{\bf A}\tau_3){\hat
  \Delta}(R),\partial_{\bf k}\rho^c_{\bf
k}\right\}\nonumber\\
&&\mbox{}\!-\!\frac{i}{8}\Big[({\bm \nabla_{\bf R}}\!-\!2ie{\bf A}\tau_3)({\bm
    \nabla_{\bf R}}\!-\!2ie{\bf A}\tau_3){\hat
  \Delta}(R),\partial_{\bf k}\partial_{\bf k}\rho^c_{\bf
k}\Big]\nonumber\\
&&\mbox{}\!-\!i\Big[\frac{1}{8m}\tau_3,{\bm \nabla}^2_{\bf R}\rho^c_{\bf
k}\Big]\!+\frac{1}{2}\!\Big\{\frac{\bf k}{m}\tau_3,{\bm \nabla}_{\bf R}\rho^c_{\bf
    k}\Big\}\nonumber\\
&&\mbox{}\!-\!e\Big[\frac{2{\bf
    A}\cdot{\bm \nabla}_{\bf R}\!+\!{\bm
      \nabla}_{\bf R}\cdot{\bf
    A}}{4m}\tau_3,\tau_3\rho^c_{\bf
    k}\Big]\!=\!\partial_t\rho^c_{\bf k}\Big|_{\rm
sc}.
\label{KE}
\end{eqnarray}
Here, $[~,~]$ and $\{~,~\}$ represent the commutator and anti-commutator,
respectively; $R=(T,{\bf R})$ denotes the center-of-mass coordinate;
$\rho^c_{\bf k}$ stands for the density matrix in the Nambu space; on the right-hand side of 
Eq.~(\ref{KE}), the scattering term $\partial_t\rho^c_{\bf k}\Big|_{\rm sc}$
is added for the completeness, whose explicit expression can be found in
Ref.~\onlinecite{GOBE4}; $\mu_H$ denotes the added gauge-invariant Hartree field, written as
\begin{equation}
\mu_H({\bf R})=\sum_{\bf R'}V_{{\bf R}-{\bf
    R'}}n({\bf R'}),\label{muH}
\end{equation}
which is equivalent to the Poisson equation. $n({\bf R})$ is the electron density.
$V_{\bf R-R'}$ denotes the Coulomb potential whose Fourier component $V_{\bf 
  q}=e^2/(q^2\epsilon_0)$.  $\epsilon_0$ represents the dielectric constant.  

The Fock energy in the pairing scheme is written as
\begin{equation}
\resizebox{.9\hsize}{!}{${\hat \Sigma}_F(R)=g{\sum_{\bf k}}'\tau_3\rho^c_{\bf k}\tau_3=\left(\begin{array}{cc}
\mu_F(R)+\mu_0 & |\Delta(R)|e^{i\theta(R)}\\
|\Delta(R)|e^{-i\theta(R)} & -\mu_F(R)+\mu_0
\end{array}\right),$}\label{Fock}
\end{equation}
where $g$ denotes the effective electron-electron attractive potential in the BCS
theory.\cite{BCS} $\sum'_{\bf k}$ here and hereafter represents the summation is restricted in the spherical shell
($|\xi_k|<\omega_D$) defined by the BCS theory.\cite{BCS} $\omega_D$ is the Debye frequency.

The effective electric field ${\bf E}$ in Eq.~(\ref{KE}),  as a gauge-invariant measurable quantity, is given by 
\begin{eqnarray}
e{\bf E}&=&-{\bm \nabla}_{\bf R}(e\phi+\mu_H+\mu_F)-\partial_Te{\bf
  A}.\label{electric}
\end{eqnarray} 

We emphasize that with the gauge structure [Eqs.~(\ref{gaugestructure1})
and~(\ref{gaugestructure2})] revealed by Nambu,\cite{gi0}
Eq.~(\ref{KE}) is gauge invariant.  In Eq.~(\ref{KE}), the third term provides the Anderson-pump 
effect. The forth and fifth terms give the drive effect. Both effects are kept here due to the gauge
invariance.\cite{GOBE1}

\subsubsection{Fock energy in GIKE}
\label{HFA}

\begin{figure}[htb]
  {\includegraphics[width=8.5cm]{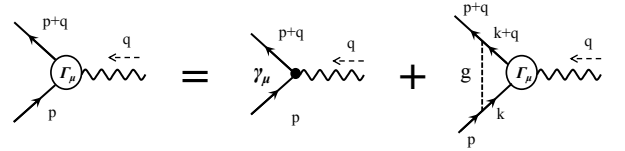}}
\caption{Diagrammatic formalism for the vertex function $\Gamma_{\mu}$. On the right-hand side of the equation, the
  first diagram corresponds to the bare vertex function (i.e., four-vector current);  the second one denotes the vertex
  correction from the pairing potential. }    
\label{figyw1}
\end{figure}

In this part, we show that the Fock energy in our GIKE approach is equivalent to the generalized Ward’s identity by
Nambu.\cite{gi0,Ba0} 
Specifically, as shown in Fig.~\ref{figyw1}, the dressed vertex function $\Gamma_{\mu}$ reads:\cite{gi0,Ba0} 
\begin{eqnarray}
\Gamma_{\mu}(p+q,p)&=&\gamma_{\mu}(p+q,p)-ig{\sum_{\bf k}}'\int\frac{dk_0}{2\pi}[\tau_3G(k+q)\nonumber\\
&&\mbox{}\times\Gamma_{\mu}(k+q,k)G(k)\tau_3], \label{Gamma}
\end{eqnarray}
in which $\gamma_{\mu}$ represents the bare
vertex function, i.e., four-vector current $\gamma_{\mu}=[\tau_3,({\bf p}+{\bf q}/2)/{m}]$;
$G(p)$ denotes $\tau_0$-Green function; $k$, $p$, $q$ are four-vector momenta. 

Substituting Eq.~(\ref{Gamma}) into the generalized Ward's identity
{\small $\sum_{\mu}q_{\mu}\Gamma_{\mu}(p+q,p)=\tau_3G^{-1}(p)-G^{-1}(p+q)\tau_3$},
one has
\begin{widetext}
\begin{eqnarray}
&&\tau_3G^{-1}(p)-G^{-1}(p+q)\tau_3=\sum_{\mu}q_{\mu}\Gamma_{\mu}(p+q,p)=\sum_{\mu}q_{\mu}\gamma_{\mu}-ig{\sum_{\bf k}}'\int\frac{dk_0}{2\pi}[\tau_3G(k+q)\sum_{\mu}q_{\mu}\Gamma_{\mu}(k+q,k)G(k)\tau_3]\nonumber\\
&&\mbox{}=-q_0\tau_3+(2{\bf p}+{\bf q})\cdot\frac{\bf q}{2m}
-ig{\sum_{\bf k}}'\int\frac{dk_0}{2\pi}[\tau_3G(k+q)-G(k)\tau_3]\nonumber\\
&&\mbox{}=\tau_3\Big[p_0-\xi_p\tau_3+ig{\sum_{\bf k}}'\int\frac{dk_0}{2\pi}\tau_3G(k)\tau_3\Big]-\Big[p_0+q_0-\xi_{\bf p+q}\tau_3+ig{\sum_{\bf k}}'\int\frac{dk_0}{2\pi}\tau_3G(k)\tau_3\Big]\tau_3.
\end{eqnarray}
\end{widetext}

Therefore, the Green function reads:
\begin{eqnarray}
G^{-1}(p)&=&p_0-\xi_p\tau_3+ig{\sum_{\bf k}}'\int\frac{dk_0}{2\pi}\tau_3G(k)\tau_3,\label{Green}
\end{eqnarray}
in which the third term on the right-hand side is the Fock energy. In a reverse way of the above derivation, one
can also prove the generalized Ward's identity by including the Fock energy in the Green function. 
Within the equal-time scheme,  the density matrix in the GIKE reads: $\rho^{c}_{\bf
  k}=-i\int{dk_0}{/(2\pi)}[\tau_3G(k)\tau_3]$.  Hence,  the Fock term in GIKE [Eq.~(\ref{Fock})]
is exactly same as that in Eq.~(\ref{Green}) above.  Therefore, the Fock energy in our GIKE approach is equivalent to the
generalized Ward's identity by Nambu.\cite{gi0,Ba0}

\subsection{Charge conservation}

In this part, facilitating with the complete Fock term,  we prove the charge conservation from the GIKE. Specifically, we first
transform Eq.~(\ref{KE}) via a unitary transformation $\rho_{\bf k}(R)={e^{-i\tau_3\theta(R)/2}}\rho^c_{\bf
  k}(R)e^{i\tau_3\theta(R)/2}$ and obtain
\begin{eqnarray}
&&\partial_T\rho_{\bf
  k}\!+\!i\left[\left(\xi_k\!+\!\mu_{\rm
      eff}\right)\tau_3\!+\!|\Delta|\tau_1,\rho_{\bf 
k}\right]\!-\!\Big[\frac{i}{8m}\tau_3,\nabla^2_{\bf R}\rho_{\bf 
k}\Big]\nonumber\\
&&\mbox{}\!-\!\frac{i}{8}\left[({\bm \nabla}_{\bf R}\!+\!i{\bf p}_s\tau_3)({\bm
    \nabla}_{\bf R}\!+\!i{\bf p}_s\tau_3)|\Delta|\tau_1,\partial_{\bf k}\partial_{\bf k}\rho_{\bf
k}\right]\nonumber\\
&&\mbox{}\!-\!\Big[\frac{2{\bf
    p}_s\cdot{\bm \nabla}_{\bf R}\!+\!{\bm \nabla}_{\bf R}\cdot{\bf p}_s}{8m}\tau_3,\tau_3\rho_{\bf
    k}\Big]\!+\!\Big\{\frac{\bf k}{2m}\tau_3,{\bm \nabla}_{\bf R}\rho_{\bf
    k}\Big\}\nonumber\\
&&\mbox{}\!+\!\frac{1}{2}\left\{e{\bf E}\tau_3\!-\!({\bm \nabla}_{\bf R}\!+\!i{\bf p}_s\tau_3)|\Delta|\tau_1,\partial_{\bf k}\rho_{\bf
k}\right\}\!=\!\partial_t\rho_{\bf k}\Big|_{\rm
sc},
\label{GE}
\end{eqnarray}
with the gauge-invariant measurable superconducting momentum ${\bf
  p}_s$ and effective field $\mu_{\rm eff}$ written as
\begin{eqnarray}
{\bf p}_s&=&\nabla_{\bf R}\theta-2e{\bf A},\label{ps}\\
\mu_{\rm eff}&=&\frac{\partial_T\theta}{2}+e\phi+\mu_{H}+\mu_{F}+\frac{p^2_s}{8m}.\label{ef}
\end{eqnarray}
By expanding the density matrix as $\rho_{\bf k}=\sum^4_{i=0}\rho_{{\bf  
    k}i}\tau_i$, each component of the Fock energy [Eq.~(\ref{Fock})] after the unitary
transformation {\small $({\hat \Sigma_F}=g\sum_{\bf k}'\tau_3\rho_{\bf k}\tau_3=\mu_0\tau_0+\mu_F\tau_3+|\Delta|\tau_1)$} reads:
\begin{eqnarray}
g{\sum_{\bf k}}'\rho_{{\bf k}3}&=&\mu_F\label{muF},\\
g{\sum_{\bf k}}'\rho_{{\bf k}1}&=&-|\Delta|\label{gap},\\
g{\sum_{\bf k}}'\rho_{{\bf k}2}&=&0\label{phase}.
\end{eqnarray}
It is noted that Eq.~(\ref{gap}) gives the gap equation, from which one can
self-consistently obtain the Higgs mode.  We show in the following section that from Eq.~(\ref{phase}), the NG mode,
which has been overlooked in our previous works,\cite{GOBE1,GOBE2,GOBE3,GOBE4}
can be self-consistently determined.

The gauge-invariant charge density $en$ and current ${\bf j}$ read:\cite{GOBE4}
\begin{eqnarray}
en&=&e\sum_{\bf k}\left[1+{\rm Tr}(\rho^c_{\bf k}\tau_3)\right]=e\sum_{\bf
  k}\left(1+2\rho_{{\bf k}3}\right), \label{density}\\  
{\bf j}&=&\sum_{\bf k}{\rm
  Tr}\left(\frac{e{\bf k}}{m}\rho^c_{\bf k}\right)=2\sum_{\bf k}\left(\frac{e{\bf k}}{m}\rho_{{\bf k}0}\right).~~~\label{current}
\end{eqnarray}
Then, from the $\tau_3$ component of the GIKE
[Eq.~(\ref{GE})]: 
\begin{eqnarray}
&&\partial_{T}\rho_{{\bf k}3}\!+\!\frac{{\bf k}\cdot{\bm \nabla}_{\bf R}\rho_{{\bf
    k}0}}{m}\!-\!2|\Delta|\rho_{{\bf k}2}\!=\!-(e{\bf E}\cdot{\partial}_{\bf k})\rho_{{\bf
    k}0}\!-\!\frac{1}{4}\nonumber\\
&&\mbox{}\!\times\!\{\partial_{\bf k}\partial_{\bf k}\!:\![\rho_{{\bf k}2}({\bm \nabla}_{\bf R}{\bm \nabla}_{\bf R}\!-\!{\bf
  p}_s{\bf p}_s)|\Delta|\!-\!\rho_{{\bf k}1}\{{\bm \nabla}_{\bf R},{\bf
  p}_s\}|\Delta|]\},\nonumber\\
\label{C1}
\end{eqnarray}
considering the fact that the right-hand side of Eq.~(\ref{C1}) vanishes after the summation of ${\bf k}$,
one has
\begin{eqnarray}
\partial_{T}\big(\sum_{\bf k}\rho_{{\bf k}3}\big)+{\bm \nabla}_{\bf R}\cdot\big(\sum_{\bf k}\frac{{\bf k}}{m}\rho_{{\bf
    k}0}\big)=2|\Delta|{\sum_{\bf k}}'\rho_{{\bf k}2},~~\label{cc}
\end{eqnarray}
in which we have used the fact that the gap vanishes outside the spherical shell in the BCS theory,\cite{G1,BCS}
$\sum_{\bf k}|\Delta|\rho_{{\bf k}2}={\sum_{\bf  k}}'|\Delta|\rho_{{\bf k}2}$. Consequently,  since the right-hand side
of Eq.~(\ref{cc}) is zero  because of Eq.~(\ref{phase}), by looking into the charge density
[Eq.~(\ref{density})] and current [Eq.~(\ref{current})] expressions, 
one immediately obtains the charge conservation:
\begin{equation}
\partial_{T}en+{\bm \nabla}_{\bf R}\cdot{\bf j}=0.
\end{equation}

Therefore, in the GIKE approach, the charge conservation is naturally satisfied with the complete Foch term
[Eqs.~(\ref{gap}) and (\ref{phase})], in contrast to the Ambegaokar and Kadanoff's approach\cite{AK}
where an additional condition of the charge conservation is required to obtain NG mode. This is because
that the Fock energy in the GIKE is equivalent to the generalized Ward's identity by Nambu,\cite{gi0,Ba0} as
proved in Sec.~\ref{HFA}, and hence, the gauge invariance in the GIKE directly leads to the charge conservation.

\section{Analytic Analysis}

In this section, we perform the analytical analysis for the electromagnetic response of
the collective Higgs and NG modes in the linear and second-order regimes.
By assuming the external electromagnetic potential $\phi=\phi_0(\omega,{\bf
  q})e^{i\omega{t}-i{\bf q}\cdot{\bf 
  R}}$ and  ${\bf A}={\bf A}_0(\omega,{\bf q})e^{i\omega{t}-i{\bf q}\cdot{\bf
  R}}$, the density matrix $\rho_{\bf k}$ and charge density $en$ read:
\begin{eqnarray}
\rho_{\bf k}&=&\rho^0_{\bf k}+\rho^{\omega}_{\bf k}e^{i\omega{t}-i{\bf q}\cdot{\bf 
  R}}+\rho^{2\omega}_{\bf k}e^{2i\omega{t}-2i{\bf q}\cdot{\bf 
  R}},\\
en&=&en_0+en^{\omega}e^{i\omega{t}-i{\bf q}\cdot{\bf 
  R}}+en^{2\omega}e^{2i\omega{t}-2i{\bf q}\cdot{\bf R}},
\end{eqnarray} 
whereas the phase $\theta$
and amplitude $|\Delta|$ of the order parameter are written as
\begin{eqnarray}
\theta&=&\theta^{\omega}e^{i\omega{t}-i{\bf q}\cdot{\bf 
  R}}+\theta^{2\omega}e^{2i\omega{t}-2i{\bf q}\cdot{\bf 
  R}},\\
|\Delta|&=&\Delta_0+\delta|\Delta|^{\omega}e^{i\omega{t}-i{\bf q}\cdot{\bf 
  R}}+\delta|\Delta|^{2\omega}e^{2i\omega{t}-2i{\bf q}\cdot{\bf 
  R}}.~~~~
\end{eqnarray} 
Here, $\rho^0_{\bf k}$, $en_0$ and $\Delta_0$ are the density matrix, charge density and order parameter in equilibrium
state, respectively; $\rho^{\omega (2\omega)}_{\bf  
  k}$, $en^{\omega (2\omega)}$, $\theta^{\omega (2\omega)}$ and
$\delta|\Delta|^{\omega (2\omega)}$ denote the linear (second-order)
responses of the density matrix, charge density, Higgs mode and NG mode, respectively.  

The density matrix in equilibrium state is given by\cite{GOBE1,GOBE4}
\begin{equation}
\rho^0_{\bf k}=\frac{1}{2}-\frac{f_k}{2}\left(\frac{\xi_k}{E_k}\tau_3+\frac{\Delta_0}{E_k}\tau_1\right)
\end{equation}
with $f_k=1-2n_F(E_k)$ and $E_k=\sqrt{\xi_k^2+\Delta_0^2}$. Here, $n_F(x)$ is
the Fermi-distribution function. From Eq.~(\ref{gap}), $\Delta_0$ is determined by 
\begin{equation}
\Delta_0=-g{\sum_{\bf k}}'\rho^0_{{\bf k}1}=g{\sum_{\bf k}}'\left(\frac{\Delta_0}{2E_k}f_k\right),\label{gap0}
\end{equation} 
which is exactly the gap equation in the BCS theory.\cite{BCS}  
$en_0$ from Eq.~(\ref{density}) is written as 
\begin{equation}
en_0=e\sum_{\bf k}(1+2\rho^0_{{\bf k}3})=\sum_{\bf k}\Big[2ev^2_k+2e\frac{\xi_k}{E_k}n_F(E_k)\Big],
\end{equation}
consisting of the charge densities of the condensate\cite{cn0,cn1,cn2,cn3}
$2ev^2_k=e(1-\frac{\xi_k}{E_k})$ and Bogoliubov quasiparticles\cite{cn0,cn1,cn2,cn3,cn4,cn5}
$2e\frac{\xi_k}{E_k}n_F(E_k)$.

Then, we show that the GIKE [Eq.~(\ref{GE})] provides an efficient approach to study the
electromagnetic  responses of the collective NG and Higgs modes. 

\subsection{Linear response}

We first focus on the linear response in this part. From Eqs.~(\ref{ps}) and (\ref{ef}), the linear 
responses of the superconducting momentum ${\bf p}_s^{\omega}$ and effective field $\mu_{\rm
  eff}^{\omega}$ are given by
\begin{eqnarray}
{\bf p}^{\omega}_s&=&-i{\bf q}\theta^{\omega}-2e{\bf A}_0,\label{ps1}\\
\mu^{\omega}_{\rm
  eff}&=&\frac{i\omega\theta^{\omega}}{2}+e\phi_0+\mu^{\omega}_{H}+\mu^{\omega}_{F}.\label{ef1}
\end{eqnarray}
with the linear responses of the Hartree field $\mu^{\omega}_H$
[Eq.~(\ref{muH})] and Fock one $\mu^{\omega}_F$ [Eq.~(\ref{muF})] written as
\begin{eqnarray}
\mu^{\omega}_H&=&V_qn^{\omega}=2V_q\sum_{\bf k}\rho^{\omega}_{{\bf k}3},\label{muH1}\\
\mu^{\omega}_F&=&g\sum_{\bf k}\rho^{\omega}_{{\bf k}3}.\label{muF1}
\end{eqnarray}
We then investigate the linear responses of the NG mode $\theta^{\omega}$ and Higgs
mode $\delta|\Delta|^{\omega}$.  

\subsubsection{NG  mode}

We address the NG mode in this part. In the long-wave limit, we only keep the lowest two 
orders of ${\bf q}$. In this situation, the linear response of the density matrix $\rho^{\omega}_{\bf k}$ can 
be solved from the GIKE. Substituting the linear solution of 
$\rho^{\omega}_{{\bf k}2}$ into Eq.~(\ref{phase}), one has (refer to Appendix~\ref{aa}) 
\begin{equation}
i\omega\mu^{\omega}_{\rm eff}\left(1+\frac{q^2v^2_F}{3\omega^2}g_{\omega}\right)+\frac{i{\bf q}\cdot{\bf
    p}_s^{\omega}}{2}\frac{v_F^2}{3}s_{\omega}=i\omega\delta|\Delta|^{\omega}b_{\omega}, \label{phase1}
\end{equation}
with the dimensionless factors:
\begin{eqnarray}
s_{\omega}&=&\frac{{\sum_{\bf
    k}}'\Big[\frac{1}{4E_k^2-\omega^2}\big(2-{E_k^2\partial^2_{\xi_k}}\big)\big(\frac{\Delta_0}{2E_k}f_k\big)\Big]}{{\sum_{\bf
    k}}'\big(\frac{1}{4E_k^2-\omega^2}\frac{\Delta_0}{E_k}f_k\big)},\label{sw}\\
g_{\omega}&=&\frac{{\sum_{\bf
    k}}'\Big[\frac{\Delta_0}{4E_k^2-\omega^2}\partial_{\xi_k}\big(\frac{\xi_k}{E_k}f_k\big)\Big]}{{\sum_{\bf
    k}}'\big(\frac{1}{4E_k^2-\omega^2}\frac{\Delta_0}{E_k}f_k\big)},\label{gw}\\
b_{\omega}&=&\frac{{\sum_{\bf
    k}}'\big(\frac{1}{4E_k^2-\omega^2}\frac{\xi_k}{E_k}f_k\big)}{{\sum_{\bf
    k}}'\big(\frac{1}{4E_k^2-\omega^2}\frac{\Delta_0}{E_k}f_k\big)}=0.\label{bw}
\end{eqnarray}
Here, we have taken care of the particle-hole symmetry to remove terms with the
odd order of $\xi_k$ in the summation of ${\bf k}$. Consequently, since
$b_{\omega}=0$ [Eq.~(\ref{bw})], it is obvious that the linear response of the NG mode
decouples with that of the Higgs mode ($\delta|\Delta|^{\omega}$) due to the
particle-hole symmetry, in consistency with the symmetry
analysis.\cite{symmetry}   

Further substituting ${\bf p}_s^{\omega}$ [Eq.~(\ref{ps1})] and
$\mu^{\omega}_{\rm eff}$ [Eq.~(\ref{ef1})] into Eq.~(\ref{phase1}), one obtains the linear-response equation of the NG mode:
\begin{widetext}
\begin{equation}
\Big[\omega^2-\frac{q^2v_F^2}{3}(s_{\omega}-g_{\omega})\Big]\frac{\theta^{\omega}}{2}=i\omega{e\phi_0}-\frac{v^2_F}{3}s_{\omega}i{\bf 
q}\cdot{e{\bf
  A}_0}-\frac{q^2v^2_F}{3}g_{\omega}\frac{e\phi_0}{i\omega}+i\omega(\mu^{\omega}_H+\mu^{\omega}_F)\Big(1+\frac{q^2v^2_F}{3\omega^2}g_{\omega}\Big).\label{phase2}
\end{equation}
\end{widetext}

We first discuss the situation without the Hartree and Fock terms. In the low-frequency regime with
$\omega\ll2\Delta_0$, one finds $s_{\omega}\approx1$ and
$g_{\omega}\approx2/3$ (refer to Appendix~\ref{aa}). Hence, the linear-response equation of
the NG mode [Eq.~(\ref{phase2})] becomes
\begin{small}
\begin{equation}
\Big[\omega^2-\Big(\frac{qv_F}{3}\Big)^2\Big]\frac{\theta^{\omega}}{2}=i\omega{e\phi_0}\Big[1+2\Big(\frac{qv_F}{3\omega}\Big)^2\Big]-\frac{v^2_F}{3}i{\bf    
q}\cdot{e{\bf
  A}_0}.\label{phase3} 
\end{equation} 
\end{small}

Consequently, it is found that the collective NG mode exhibits the gapless
linear energy spectrum (i.e., $\omega_{\rm NG}={qv_F}/{3}$) inside the
superconducting gap, in consistency with the previous
works\cite{gi0,AK,Ba0,pm0,Am0,pi1,Ba9,Ba10,gi1,pm5} and Goldstone theorem with the
spontaneous continuous $U(1)$-symmetry breaking.\cite{Gm1,Gm2}      
Additionally, the NG mode responds to the longitudinal electromagnetic field [right-hand side of 
Eq.~(\ref{phase3})] in the linear regime, also in agreement with the previous works.\cite{AK,Ba0,Ba9,Ba10}  

\subsubsection{Role of Hartree and Fock fields}
\label{RHF}

We next consider the role of the Hartree and Fock fields in the linear response of the NG mode. 
Specifically, considering $V_q\gg{g}$ in the long-wave limit, the
Fock field can be neglected. Substituting the solution of $\rho^{\omega}_{{\bf k}3}$ into Eq.~(\ref{muH1}), the
Hartree field reads (refer 
to Appendix~\ref{Eff}):  
\begin{equation}
\mu^{\omega}_H=\frac{V_q{\bf q}\cdot{\bf
    E}^{\omega}}{im\omega^2}{\sum_{\bf
  k}}'\frac{k_F^2}{3m}\Big[{\partial_{E_k}f_k}-\frac{\Delta^2_0}{E_k}\partial_{E_k}\Big(\frac{f_k}{E_k}\Big)\Big].\label{plasma}
\end{equation}
It is noted that the first term in the summation of ${\bf k}$ denotes
the contribution from the Bogoliubov quasiparticles. The second one exactly corresponds to the Meissner-superfluid
density {\small $\rho_s=\sum_{\bf
    k}'\frac{k_F^2}{3m}\big[\frac{\Delta^2_0}{E_k}\partial_{E_k}\big(-\frac{f_k}{E_k}\big)\big]$}, related to the
Meissner supercurrent, as revealed in our
previous work.\cite{GOBE4}

At low temperature, the Bogoliubov quasiparticles vanish, i.e., $n_F(E_k)\approx0$, leaving solely the Meissner-superfluid
density. Then, one has $\mu^{\omega}_H=-\frac{i{\bf 
    q}\cdot{e}{\bf E}^{\omega}}{{q^2}}\frac{\omega^2_p}{\omega^2}$
with $\omega_p=\sqrt{\frac{\rho_se^2}{\epsilon_0m}}$ being the plasma frequency.
Further substituting ${\bf E}^{\omega}$ [Eq.~(\ref{electric})] into $\mu^{\omega}_H$, the Hartree field
is given by 
\begin{equation}
\mu^{\omega}_H=-\frac{i{\bf q}{\cdot}e{\bf
    E}^{\omega}_0}{q^2}\frac{\omega^2_p/\omega^2}{1-{\omega^2_p}/{\omega^2}}, \label{muHFF}
\end{equation}
with $e{\bf E}_0=i{\bf q}\phi_0-i\omega{\bf A}_0$ being the external electric field.

Finally, considering the contribution of the Hartree field [Eq.~(\ref{muHFF})],
the linear-response equation [Eq.~(\ref{phase2})] of the NG mode becomes
\begin{eqnarray}
\label{phase4}
\Big[\omega^2-\Big(\frac{qv_F}{3}\Big)^2\Big]\frac{\theta^{\omega}}{2}=\frac{i\omega{e\phi_0-\omega^2_p{i{\bf
    q}{\cdot}e{\bf A}_0}/{q^2}}}{1-\omega^2_p/\omega^2}+O(q).~~
\end{eqnarray} 
Therefore, as seen from the right-hand side of Eq.~(\ref{phase4}), as a consequence of the Hartree field (i.e., the vacuum
polarization), the longitudinal field experiences the Coulomb screening. In this situation, multiplying by
$1-\omega^2_p/\omega^2$ on both sides of Eq.~(\ref{phase4}), in the long-wave limit, one has  
\begin{equation}
(\omega^2-\omega_p^2)\frac{\theta^{\omega}}{2}=i\omega{e\phi_0}-\omega^2_p\frac{i{\bf
    q}{\cdot}e{\bf A}_0}{q^2},\label{phase5}
\end{equation}
exactly same as the previous work.\cite{AK}  Consequently, as seen from Eq.~(\ref{phase5}), 
the linear response of the NG mode interacts with the long-range Coulomb interaction, causing the original gapless
spectrum of the NG mode {\em effectively} lifted up to the high-energy plasma frequency as a result of the
Anderson-Higgs mechanism.\cite{AK,Ba0,pm0,pm5,Am0,AHM} 

With the high-energy plasma frequency (i.e.,  $\omega\ll\omega_p$), one finds ${\theta^{\omega}}/{2}={i{\bf 
    q}{\cdot}e{\bf 
    A}_0}/{q^2}$.  As pointed out in the previous works,\cite{Ba0,AK}  this finite $\theta^{\omega}$ from the
    unphysical longitudinal vector potential does not provide any measurable effect, especially considering the fact
    that the longitudinal vector potential does not even exist in either optical response or static magnetic response.
    Moreover, this finite $\theta^{\omega}$ cancels the unphysical longitudinal vector
    potential in ${\bf p}^{\omega}_s$ [Eq.~(\ref{ps1})]: 
    \begin{equation}
      \frac{{\bf p}^{\omega}_s}{2}=\frac{{\bf q}({\bf q}{\cdot}e{\bf
          A}_0)}{q^2}-e{\bf A}_0+O\Big(\frac{\omega^2}{\omega^2_p}\Big).\label{psf}
    \end{equation}
    As a result,  the gauge-invariant superconducting momentum ${\bf p}^{\omega}_s$, which appears in the
    Ginzburg-Landau equation,\cite{G1,GOBE4} Meissner supercurrent\cite{G1,GOBE4} and 
    Anderson-pump effect,\cite{Am1,Am2,Am7,Am9,Am11,Am14,NL7,NL8,NL9,NL10,NL11,Am3,Am4,Am8,Am10,GOBE1,GOBE2,GOBE3,GOBE4}   only involves the physical transverse vector potential.

Interestingly, at low temperature, it is observed above that the emerged plasma
frequency $\omega_p=\sqrt{\frac{\rho_se^2}{\epsilon_0m}}$ origins from the
  Meissner-superfluid density $\rho_{s}=\sum_{\bf 
  k}'\frac{k_F^2}{3m}\frac{\Delta_0^2}{E^3_k}$, rather than the condensate $en_0=\sum_{\bf
  k}2ev_k^2$. This is in consistency with our previous conclusion\cite{GOBE4} that only the
Meissner-superfluid density, which is related to the charge fluctuation on top of the
condensate, is involved in the electromagnetic response in the superconducting
states whereas the ground state condensate simply provides a rigid background.

\subsubsection{Higgs mode}

We next study the linear response of the Higgs mode. Substituting the second-order solution of  $\rho^{\omega}_{{\bf
    k}1}$ into  the gap equation [Eq.~(\ref{gap})], one directly obtains (refer to Appendix~\ref{aa}) 
\begin{eqnarray}
&&{i\omega\delta|\Delta|^{\omega}}\Big[\frac{1}{g}-{\sum_{\bf
k}}'\Big(\frac{2\xi^2_k}{4E^2_k-\omega^2}\frac{f_k}{E_k}\Big)\Big]=-\frac{i({\bf q}\cdot{\bf
  p}^{\omega}_s)c_{\omega}}{6m}\nonumber\\
&&\mbox{}+\frac{i{\bf q}}{m}\cdot\frac{e{\bf
  E}^{\omega}}{i\omega}{\sum_{\bf
k}}'\Big[\frac{4\xi^2_k}{4E^2_k-\omega^2}\frac{2}{3}\partial_{\xi_k}(\xi_k\rho^0_{{\bf
  k}1})\Big],\label{higgs1}
\end{eqnarray}
where the particle-hole symmetry has been taken care of to remove terms with 
odd order of $\xi_k$ in the summation of ${\bf k}$. $c_{\omega}$ is a dimensionless
factor (refer to Appendix~\ref{aa}).

The first term on the right-hand side of Eq.~(\ref{higgs1}) vanishes since ${\bf p}^{\omega}_s$ only involves the
physical transverse vector potential [Eq.~(\ref{psf})].  By using Eq.~(\ref{gap0}) 
to replace $g$, the linear response of the Higgs mode is obtained:
\begin{equation}
i\omega\delta|\Delta|^{\omega}\Big[1-\Big(\frac{\omega}{2\Delta_0}\Big)^2\Big]=u_{\omega}\frac{i{\bf q}{\cdot}e{\bf
  E}^{\omega}}{im\omega},\label{higgs2}
\end{equation}
with {\small $u_{\omega}={{\sum_{\bf
k}}'\Big[\frac{4\xi^2_k\partial_{\xi_k}(\xi_k\rho^0_{{\bf
  k}1})}{4E^2_k-\omega^2}\Big]}/{{\sum_{\bf
k}}'\Big(\frac{3\Delta_0^2}{4E^2_k-\omega^2}\frac{f_k}{E_k}\Big)}$}.

Consequently, from Eq.~(\ref{higgs2}), it is seen that the Higgs mode exhibits the gapful energy spectrum (i.e.,
$\omega_{\rm 
  H}=2\Delta_0$), in consistency with the previous studies.\cite{Am0,pm5} Moreover, the
Higgs mode also responds to the electromagnetic field in the linear 
regime [right-hand side of Eq.~(\ref{higgs2})]. Nevertheless, this linear response vanishes in the long-wave limit,
making it hard to be detected in the optical experiment.    

\subsection{Second-order response}

From above analytic investigations, one directly concludes that neither the collective phase (NG) mode nor the amplitude (Higgs) 
mode is detectable in the linear regime for the optical experiment. In contrast,  we show in this section that
the second-order response of the collective modes in superconductors exhibits different physics.   

Specifically, the second-order responses of the superconducting momentum ${\bf 
  p}_s^{2\omega}$ and effective field $\mu_{\rm eff}^{2\omega}$ from
Eqs.~(\ref{ps}) and (\ref{ef}) are given by
\begin{eqnarray}
{\bf p}^{2\omega}_s&=&-2i{\bf q}\theta^{2\omega},\label{ps2}\\
\mu^{2\omega}_{\rm eff}&=&i\omega\theta^{2\omega}+\mu^{2\omega}_{H}+\mu^{2\omega}_{F}+\frac{(p^{\omega}_s)^2}{8m}.\label{ef2}
\end{eqnarray}
It is noted that the last term on the right-hand side of Eq.~(\ref{ef2}) is
exactly the Anderson-pump effect.
 
The second-order responses of the Hartree field $\mu^{2\omega}_H$
[Eq.~(\ref{muH})] and Fock one $\mu^{2\omega}_F$ [Eq.~(\ref{muF})] are written as
\begin{eqnarray}
\mu^{2\omega}_H&=&V_{2q}n^{2\omega}=2V_{2q}\sum_{\bf k}\rho^{2\omega}_{{\bf k}3},\label{muH2}\\
\mu^{2\omega}_F&=&g\sum_{\bf k}\rho^{2\omega}_{{\bf k}3}.\label{muF2}
\end{eqnarray}
Then, we investigate the second-order responses of the NG mode $\theta^{2\omega}$ and Higgs
mode $\delta|\Delta|^{2\omega}$.  

\subsubsection{NG mode}
\label{NG2}

We address the NG mode in this part. The second-order response of the density
matrix $\rho^{2\omega}_{\bf k}$ can 
also be obtained from the GIKE in the long-wave
limit. Substituting the solution of $\rho^{2\omega}_{{\bf k}2}$ into 
Eq.~(\ref{phase}), one has  (refer to Appendix~\ref{bb})  
\begin{eqnarray}
&&2i\omega\mu^{2\omega}_{\rm eff}\Big(1+\frac{q^2v^2_F}{3\omega^2}g_{2\omega}\Big)+{i{\bf q}\cdot{\bf
  p}_s^{2\omega}}\frac{v^2_F}{3}s_{2\omega}=\frac{2i\omega}{m}\Big[\frac{g_{2\omega}}{3}\nonumber\\
&&\mbox{}\times\Big(\frac{e{\bf
      E}^{\omega}}{i\omega}-{{\bf p}^{\omega}_s}\Big){\cdot}\frac{e{\bf
  E}^{\omega}}{i\omega}+\frac{l_{\omega}}{2}\Big(\frac{e{\bf E}^{\omega}}{i\omega}-\frac{{\bf
      p}^{\omega}_s}{2}\Big)^2\Big],~~~~~~\label{sphase1}
\end{eqnarray}
with dimensionless prefactor:
\begin{equation}
l_{\omega}=\frac{{\sum_{\bf k}}'\Big[\frac{\Delta_0}{E_k^2-\omega^2}(2\xi_k\partial^2_{\xi_k}+\partial_{\xi_k})\Big(\frac{\xi_k}{E_k}f_k\Big)\Big]}{3{\sum_{\bf
  k}}'\Big(\frac{1}{E_k^2-\omega^2}\frac{\Delta_0}{E_k}f_k\Big)}. \label{lw}
\end{equation}

Furthermore, with the solution of $\rho^{2\omega}_{{\bf k}3}$, we find that the
second-order response of the charge density $en^{2\omega}=e\sum_{\bf
  k}2\rho^{2\omega}_{{\bf k}3}$ is zero (refer to Appendix~\ref{bb}), leading to the
vanishing second-order Hartree field $\mu^{2\omega}_H$ [Eq.~(\ref{muH2})] and 
Fock one $\mu^{2\omega}_F$ [Eq.~(\ref{muF2})]. 

Consequently, substituting
${\bf p}_s^{2\omega}$ [Eq.~(\ref{ps2})] and $\mu^{2\omega}_{\rm eff}$ [Eq.~(\ref{ef2})] into
Eq.~(\ref{sphase1}), the second-order response equation of the NG mode
reads:
\begin{eqnarray}
&&\Big[\omega^2-\frac{q^2v_F^2}{3}(s_{2\omega}-g_{2\omega})\Big]\theta^{2\omega}=\frac{i\omega}{m}\Big[\frac{({\bf
    p}_s^{\omega})^2}{8}-\frac{g_{2\omega}}{3}\nonumber\\
&&\mbox{}\times\Big(\frac{e{\bf
      E}^{\omega}}{i\omega}-{{\bf p}^{\omega}_s}\Big){\cdot}\frac{e{\bf
  E}^{\omega}}{i\omega}-\frac{l_{\omega}}{2}\Big(\frac{e{\bf E}^{\omega}}{i\omega}-\frac{{\bf
      p}^{\omega}_s}{2}\Big)^2\Big],~~~~~~\label{sphase2}
\end{eqnarray}
which exhibits different physics from the linear response. 

Particularly,  in the low-frequency regime ($\omega\ll{\Delta_0}$), one finds
that $s_{2\omega}\approx1/3$, $g_{2\omega}\approx2/3$, (refer to Appendix~\ref{aa}) and $l_{\omega}\approx-2/45$ 
(refer to Appendix~\ref{bb}), and hence, Eq.~(\ref{sphase2}) becomes
\begin{equation}
\Big(\omega^2-\frac{q^2v_F^2}{9}\Big){\theta^{2\omega}}\approx\frac{i\omega}{m}\frac{({\bf
    p}_s^{\omega})^2}{8}+\Big({{\bf p}^{\omega}_s}-\frac{e{\bf
      E}^{\omega}}{i\omega}\Big){\cdot}\frac{e{\bf
  E}^{\omega}}{5m}.\label{sphase3}
\end{equation}
On the right-hand side of Eq.~(\ref{sphase3}), the first term exactly comes from the Anderson pump effect whereas the
last two ones are attributed to the second order of the drive effect. Both effects play important role in the
second-order response of the NG mode. Moreover, it is noted that on the right-hand side of in Eq.~(\ref{sphase2}),
${\bf p}^{\omega}_s$ only involves the physical transverse vector potential [Eq.~(\ref{psf})]. As for the electric field
${\bf E}^{\omega}={\bf E}^{\omega,\parallel}+{\bf E}^{\omega,\perp}$, by the linear response of the Hartree field
$\mu^{\omega}_H$ (i.e., the vacuum polarization), the longitudinal electric field  ${\bf E}^{\omega,\parallel}$ is
suppressed by the strong Coulomb screening whereas the transverse one  ${\bf E}^{\omega,\perp}$ is not affected (refer
to Appendix~\ref{Eff}). Therefore, the second-order response of the NG mode at low frequency ($\omega{\ll}\omega_p$) 
is determined by the transverse field.

Consequently, from Eq.~(\ref{sphase3}), it is very interesting to find that due to the vanishing Hartree field,  the
second-order response of the NG mode maintains the original gapless energy spectrum ($\omega_{\rm NG}=qv_F/3$) inside
the superconducting gap, in striking contrast to the linear response with the Anderson-Higgs mechanism.  
This can be understood as follows.  In the presence of the inverse symmetry, no second-order current is induced,
and hence, due to the charge conservation, no charge density fluctuation $en^{2\omega}$ is excited, effectively ruling
out the Hartree field $\mu^{2\omega}_H=V_qn^{2\omega}$ (i.e., the long-range Coulomb interaction) in the second-order
response. In  addition, differing from the linear response excited by the longitudinal field solely,  the second-order
response of the  NG mode is determined by the transverse field as mentioned above,  free from the influence of the
Coulomb screening.   

We point out that thanks to the gauge-invariant electric field and superconducting moment on the right-hand side of
Eq.~(\ref{sphase3}), the second-order response of the NG mode $\theta^{2\omega}$ is a measurable quantity, differing
from the linear response above. This term, which is hard to deal with in the previous
approaches,\cite{AK,Ba0,Ba9,Ba10,pm0,pm5,Am0,pi1,pi4} has long been overlooked in the literature.

\subsubsection{Higgs mode}
\label{Higgs2}

Substituting the solution of $\rho^{2\omega}_{{\bf k}1}$ into Eq.~(\ref{gap}), in the long-wave
limit, one has (refer to Appendix~\ref{bb})
\begin{eqnarray}
\frac{\delta|\Delta|^{2\omega}}{g}\!=\!{\sum_{\bf
  k}}'\frac{\xi_k\Big[\delta|\Delta|^{2\omega}+\big(\frac{{\bf p}^{\omega}_s}{2}-\frac{e{\bf
      E}^{\omega}}{i\omega}\big)^2\frac{\Delta_0v^2_F\partial^2_{\xi_k}}{6}\Big]\rho^0_{{\bf
    k}3}}{\omega^2-E^2_k}.~~\label{shiggs2}
\end{eqnarray}
By using Eq.~(\ref{gap0}) to replace $g$, the second-order response equation of the Higgs mode in the long-wave
limit reads:
\begin{equation}
\delta|\Delta|^{2\omega}\Big[1-\Big(\frac{2\omega}{2\Delta_0}\Big)^2\Big]=\frac{v_F^2}{6}\Big(\frac{{\bf
      p}^{\omega}_s}{2}-\frac{e{\bf E}^{\omega}}{i\omega}\Big)^2\frac{d_{\omega}}{\Delta_0},\label{shiggs1}
\end{equation}
with {\small $d_{\omega}={{\sum_{\bf
    k}}'\Big[\frac{\xi_k\Delta_0}{E_k^2-\omega^2}\partial^2_{\xi_k}\Big(\frac{\xi_k}{E_k}f_k\Big)\Big]}/{{\sum_{\bf
  k}}'\Big(\frac{f_k}{E_k}\frac{\Delta_0}{E_k^2-\omega^2}\Big)}$}. 

Therefore, a finite response of the Higgs mode in the long-wave limit is found in the second-order regime,  
differing from the vanishing linear response above. Furthermore, this second-order response of the Higgs mode,
shows a resonance at $2\omega=2\Delta_0$, in consistency with the experimental findings.\cite{NL8,NL9,NL10}   
Particularly, we point out that the right-hand side of Eq.~(\ref{shiggs1}) exactly comes from the second-order of drive
effect whereas the widely considered pump effect in the
literature\cite{Am1,Am2,Am7,Am9,Am11,Am14,NL7,NL8,NL9,NL10,NL11,Am3,Am4,Am8,Am10,GOBE1,GOBE2,GOBE3,GOBE4} makes no  contribution at all. 

Actually, it is noted that in the previous theoretical
studies,\cite{Am1,Am2,Am3,Am4,Am5,Am6,Am7,Am8,Am9,Am10,Am11,Am12,Am13,Am14,NL7,NL8,NL9,NL10,NL11,
  GOBE1,GOBE2,GOBE3,GOBE4} the obtained fluctuation of the order parameter $\delta\Delta^{2\omega}$ is 
directly considered as the amplitude (Higgs) mode $\delta|\Delta|^{2\omega}$ since it is believed that 
the phase (NG) mode is lifted up to the high-energy plasma frequency. Then,  it is considered that 
the Anderson-pump effect, which can excite the fluctuation of the order parameter $\delta\Delta^{2\omega}$, contributes
to the amplitude mode.   Nevertheless,  this becomes ambiguous when the very recent symmetry analysis by
Tsuchiya {\em et al.}\cite{symmetry} implies  that the pump effect excites the oscillation of the superconducting phase
rather than the amplitude.  Even though not clearly stated, the obtained pseudospin susceptibilities $\chi_{yz}\ne0$ and 
$\chi_{xz}=0$ [Eq.~(25) in Ref.~\onlinecite{symmetry}] in their work clearly suggest that  the induced
pseudo field $H_z$ by the pump effect in the Anderson pseudospin
picture\cite{Am1,Am2,Am7,Am9,Am11,Am14,NL7,NL8,NL9,NL10,NL11} can only generate the fluctuation of the phase-related
$S_y$, rather than the amplitude-related $S_x$.  To resolve this puzzle, the 
contributions from the amplitude and phase modes to $\delta\Delta^{2\omega}$ in the previous works\cite{Am1,Am2,Am3,Am4,Am5,Am6,Am7,Am8,Am9,Am10,Am11,Am12,Am13,Am14,NL7,NL8,NL9,NL10,NL11,GOBE1,GOBE2,GOBE3,GOBE4}
must be carefully examined.     

In contrast, the GIKE provides an efficient approach to calculate the phase and amplitude modes on an
equal footing.  The results from the GIKE above suggest that the fluctuation of the order parameter in
the second-order response actually consists of contributions from both amplitude (Higgs) and phase (NG) modes, i.e.,
$\delta\Delta^{2\omega}=\delta|\Delta|^{2\omega}+i\theta^{2\omega}\Delta_0$.  From above analytic analysis, we
conclude that the observed second-order response of the amplitude mode $\delta|\Delta|^{2\omega}$ in the recent 
optical experiments\cite{NL7,NL8,NL9,NL10,NL11} is attributed solely to the drive effect rather than the widely
considered Anderson-pump
effect.\cite{Am1,Am2,Am7,Am9,Am11,Am14,NL7,NL8,NL9,NL10,NL11,Am3,Am4,Am8,Am10,GOBE1,GOBE2,GOBE3,GOBE4} 

In fact, the pump  effect only contributes to the second-order response of the NG mode $\theta^{2\omega}$, in which the
drive effect also plays an important role,  as mentioned in Sec.~\ref{NG2}.  Consequently,  all previous studies of the
Anderson-pump effect in the
literature\cite{Am1,Am2,Am3,Am4,Am5,Am6,Am7,Am8,Am9,Am10,Am11,Am12,Am13,Am14,NL7,NL8,NL9,NL10,NL11,
  GOBE1,GOBE2,GOBE3,GOBE4}   actually calculate only one part of the second-order response of the NG 
mode $\theta^{2\omega}$ rather than the Higgs mode,   supporting the latest symmetry
analysis by Tsuchiya {\em et al.}\cite{symmetry} from the Anderson pseudospin picture. Particularly,  we have
revealed in Sec.~\ref{NG2} that the second-order response of the NG mode decouples with the long-range Coulomb
interaction, free from the influence of  the Anderson-Higgs mechanism,\cite{AHM} and is measurable. A
tentative scheme to detect this second-order response of the NG mode is proposed in the following section.  

\subsubsection{Tentative scheme for detection}

\begin{figure}[htb]
  {\includegraphics[width=8.5cm]{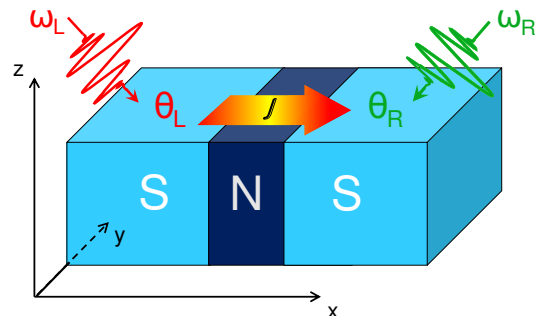}}
\caption{Schematic to detect the second-order response of the phase mode.  Two continuous-wave optical 
  fields with frequencies $\omega_L$ and $\omega_R=\omega_L/2$ are applied to the 
  superconductors on the two side of junction, leading to the excited phase $\theta_{L}$
  ($\theta_{R}$) of the left (right) superconductor.  Then,  by the
  phase difference $\theta_d=\theta_L-\theta_R$, the Josephson current $J=J_c\sin\theta_d$ is generated. }    
    
\label{figyw2}
\end{figure}

We propose a tentative scheme to detect the second-order response $\theta^{2\omega}$ through the
Josephson junction.  Specifically,  for the optical experiment, in the long-wave limit, the second-order response of the
NG mode from Eq.~(\ref{sphase3}) shows a spatially uniform but temporally oscillating phase $\theta=z_{\omega}e^{2i\omega{t}}$,
with $z_{\omega}=|\theta^{2\omega}(q=0)|$ denoting the oscillating 
amplitude of $\theta$.  Therefore, as schematically illustrated
 in Fig.~\ref{figyw2}, in a Josephson junction, 
by separately applying two phase-locked 
continuous-wave optical fields 
with frequencies $\omega_L$ and $\omega_R$ ($\omega_L=2\omega_R$) to the
superconductors on each side of junction, an oscillating phase 
difference $\theta_d=\theta_L-\theta_R$ between the
left  and right superconductors is induced, leading to the Josephson current $J=J_c\sin\theta_d$.\cite{Josephson} 
Here, $J_c$ is the Josephson critical current. Moreover, through the optical time delay to choose $\pi/2$ phase
difference, one has the phase excitations with $\theta_{L}=z^{L}_{\omega_{L}}\cos(2\omega_{L}t)$
and $\theta_{R}=z^{R}_{\omega_{R}}\sin(2\omega_{R}t)$, and then, a dc-current component in $J$ is derived
(refer Appendix~\ref{ee}):
\begin{eqnarray}
J^{\rm dc}\!=\!2J_c\!j_{1}(z^L_{\omega_{L}})j_{2}(z^R_{\omega_{R}}),\label{TD5} 
\end{eqnarray}
with $j_n(x)$ being the $n$-th Bessel function of the first kind.

Consequently, a dc current is induced. Therefore, this
dc Josephson current provides a tentative scheme for the detection of the second-order response of the NG (phase) mode,
especially  considering the fact that  the generation 
of the Josephson current directly implies the phase fluctuation.  Moreover, to avoid influence from the
optical currents, one can choose the directions of the propagation and polarization of  the applied optical fields to
be perpendicular to that of the junction, i.e., 
along $z$ and $y$ directions in Fig.~\ref{figyw2}, respectively.   

\section{SUMMARY AND DISCUSSION}
\label{summary}

We have shown that the GIKE provides an efficient approach to study the electromagnetic response of the  collective
modes in the superconducting states. We prove that the Fock energy is equivalent to the generalized Ward's 
identity by Nambu.\cite{gi0,Ba0} Therefore,  with the complete Fock term, the gauge invariance in the GIKE
directly leads to the charge conservation, in contrast to the previous Ambegaokar and Kadanoff's 
approach\cite{AK} where an additional condition of the charge conservation is required to obtain the NG mode. 
Differing from the previous studies in the literature with either the fixed amplitude\cite{AK,pi1,pm1,pm2,pi4,Ba9}
or overlooked 
phase\cite{Am1,Am2,Am3,Am4,Am5,Am6,Am7,Am8,Am9,Am10,Am11,Am12,Am13,Am14,NL7,NL8,NL9,NL10,NL11,GOBE1,GOBE2,GOBE3,GOBE4} 
of the order parameter, in the present work, the gapless NG and gapful Higgs modes are calculated on an equal footing.
Moreover, both linear and nonlinear responses are investigated. 

In the linear regime, we find that the Higgs mode responds to the electromagnetic field but this linear response vanishes in the long-wave limit. As for the NG mode, the results in the linear response by the
GIKE agree with the previous ones in the literature.\cite{AK,Ba0,pm0,pm5,Am0,Ba9,Ba10} Specifically, the linear
response of the NG mode interacts with the long-range Coulomb interaction, causing the 
original gapless spectrum inside the superconducting gap {\em effectively} lifted up to the plasma frequency far above
the gap as a result of the Anderson-Higgs mechanism.\cite{AHM} Consequently, no effective linear response of the NG mode
occurs. In addition, we reveal that the emerged plasma frequency at low temperature origins from the
Meissner-superfluid density rather than the condensate,  in consistency with our previous conclusion\cite{GOBE4} that
only the Meissner-superfluid density is involved in the electromagnetic response in the superconducting states whereas
the ground state condensate simply provides a rigid background.  Therefore, neither the collective Higgs mode nor the
NG mode is detectable in the linear regime for the optical experiment.        

The second-order responses of both collective modes exhibit interesting physics in contrast to the linear ones.
Specifically, in the second-order regime, a finite response of the Higgs mode is obtained 
in the long-wave limit.  By looking into the source of the field, we find that the widely considered Anderson-pump
effect makes no contribution at all.  Instead, only the drive effect 
contributes. Particularly, this finite second-order response  of the Higgs mode from the drive effect 
exhibits a resonance at $2\omega=2\Delta_0$, in consistency with the  experimental findings.\cite{NL8,NL9,NL10}  
Consequently, the experimentally observed second-order response of the Higgs
mode\cite{NL7,NL8,NL9,NL10,NL11} is attributed solely to the  drive effect rather than the Anderson-pump
effect widely speculated in the
literature.\cite{Am1,Am2,Am7,Am9,Am11,Am14,NL7,NL8,NL9,NL10,NL11,Am3,Am4,Am8,Am10,GOBE1,GOBE2,GOBE3,GOBE4}   

In fact, we find that the Anderson-pump effect only contributes to the second-order response of the NG mode,
in which the drive effect also plays an important role.  In addition,  we further point out that  in striking contrast to
the linear response, the second-order 
response of the NG mode  decouples with the long-range Coulomb interaction, and hence maintains the original gapless
energy spectrum inside the superconducting gap, free from the influence of  the Anderson-Higgs mechanism.\cite{AHM} 
The origin of this decoupling can be understood as follows. On one hand,
in the presence of the inverse symmetry,  no second-order current is induced in the long-wave limit. Hence,  due to
the charge conservation, no charge density fluctuation is 
excited in the second-order response, ruling out the influence of the Poisson equation (i.e., the long-range Coulomb
interaction). On the other hand,  differing from the linear response excited by the longitudinal field solely,
we find that  the second-order response of the NG mode at low frequency ($\omega{\ll}\omega_p$)  is determined by the
transverse field,  free from  the influence of the Coulomb screening.  This second-order response,  hard to completely
deal with in the previous approaches,\cite{AK,Ba0,Ba9,Ba10,pm0,pm5,Am0,pi1,pi4} has long been overlooked in the
literature.   A tentative scheme to detect this second-order response of the NG mode is proposed.

\begin{acknowledgments}
This work was supported by the National Natural Science Foundation of 
China under Grants No.\ 11334014 and No.\ 61411136001.  
\end{acknowledgments}

\begin{appendix}

\section{Derivation of Eqs.~(\ref{phase1}) and (\ref{higgs1})}
\label{aa}

In this section, we derive Eqs.~(\ref{phase1}) and (\ref{higgs1}). Considering the long-wave limit, we only keep the
lowest two orders of ${\bf q}$ in our derivation. Then, the linear order of the GIKE
[Eq.~(\ref{GE})] in the clean limit reads: 
\begin{eqnarray}
&&i\omega\rho^{\omega}_{\bf k}+i\left[\xi_k\tau_3+\Delta_0\tau_1,\rho^{\omega}_{\bf
  k}\right]-\frac{1}{2}\Big\{\frac{i{\bf k}\cdot{\bf q}}{m}\tau_3,\rho^{\omega}_{\bf
    k}\Big\}\nonumber\\
&&\mbox{}+i\left[\delta|\Delta|^{\omega}\tau_1+\mu^{\omega}_{\rm eff}\tau_3,\rho^0_{\bf
    k}\right]-\frac{i}{8}[{\bf q}{\bf
  p}^{\omega}_s\Delta_0\tau_3\tau_1,\partial_{\bf k}\partial_{\bf k}\rho^0_{\bf k}]\nonumber\\
&&\mbox{}+\frac{1}{2}\left\{e{\bf E}^{\omega}\tau_3+(i{\bf q}\delta|\Delta|^{\omega}\tau_1-i{\bf
  p}_s^{\omega}\tau_3\Delta_0\tau_1),\partial_{\bf k}\rho^0_{\bf
  k}\right\}\nonumber\\
&&\mbox{}-i\Big[\frac{{\bf q}\cdot{\bf p}_s^{\omega}}{8m}\tau_3,\tau_3\rho^0_{\bf k}\Big]=0,
\end{eqnarray}
whose components are written as
\begin{small}
\begin{eqnarray}
&&i\omega\rho^{\omega}_{{\bf k}0}=\frac{i{\bf k}\cdot{\bf
    q}}{m}\rho^{\omega}_{{\bf k}3}-{\partial_{\bf k}}\cdot(i{\bf q}\delta|\Delta|^{\omega}\rho^0_{{\bf
    k}1}+e{\bf E}^{\omega}\rho^0_{{\bf
    k}3}),~~~~~~\label{a1}\\
&&i\omega\rho^{\omega}_{{\bf k}3}-2\Delta_0\rho^{\omega}_{{\bf
    k}2}=\frac{i{\bf k}\cdot{\bf
    q}}{m}\rho^{\omega}_{{\bf k}0}+\frac{i\Delta_0}{4}{\bf q}{\bf
  p}_s^{\omega}:\partial_{\bf k}\partial_{\bf k}\rho^0_{{\bf k}1},\label{a2}~~~~~~\\
&&i\omega\rho^{\omega}_{{\bf k}1}+2\xi_k\rho^{\omega}_{{\bf
    k}2}=\frac{i{\bf
    q}\cdot{\bf p}^{\omega}_s}{4m}\rho^{0}_{{\bf k}1}-\frac{i\Delta_0}{4}{\bf q}{\bf
  p}_s^{\omega}:\partial_{\bf k}\partial_{\bf k}\rho^0_{{\bf k}3},\label{a3}\\
&&i\omega\rho^{\omega}_{{\bf k}2}+2\Delta_0\rho^{\omega}_{{\bf k}3}-2\xi_k\rho^{\omega}_{{\bf
    k}1}=({\xi_k}\delta|\Delta|^{\omega}-{\Delta_0}\mu^{\omega}_{\rm
  eff})\frac{f_k}{E_k}.~~~~~~\label{a4}
\end{eqnarray} 
\end{small}

Substituting $\rho^{\omega}_{{\bf k}3}$ [Eq.~(\ref{a2})] and $\rho^{\omega}_{{\bf k}1}$ [Eq.~(\ref{a3})] 
into Eq.~(\ref{a4}), one has 
\begin{eqnarray}
&&(4E^2_k-\omega^2)\rho^{\omega}_{{\bf k}2}=i\omega({\xi_k}\delta|\Delta|^{\omega}-{\Delta_0}\mu^{\omega}_{\rm
  eff})\frac{f_k}{E_k}+\frac{i{\bf q}\cdot{\bf p}^{\omega}_s}{2m}\nonumber\\
&&\mbox{}\times\xi_k\rho^0_{{\bf k}1}-\frac{i\Delta^2_0}{2}{\bf q}{\bf
  p}_s^{\omega}:\partial_{\bf k}\partial_{\bf k}\rho^0_{{\bf k}1}-\frac{i\xi_k\Delta_0}{2}{\bf q}{\bf
  p}_s^{\omega}:\partial_{\bf k}\partial_{\bf k}\rho^0_{{\bf k}3}\nonumber\\
&&\mbox{}+2\Delta_0\frac{i{\bf k}\cdot{\bf
    q}}{m}\frac{e{\bf E}^{\omega}\cdot\partial_{\bf k}}{i\omega}\rho^0_{{\bf k}3}+O(q^2),~~~~~~\label{ro2}
\end{eqnarray}
in which Eq.~(\ref{a1}) is used for $\rho^{\omega}_{{\bf
    k}0}$. Considering the fact: 
\begin{eqnarray}
&&\Delta_0\partial^2_k\rho^0_{{\bf
    k}3}=\frac{\rho_{{\bf k}1}^0}{m}+\xi_k\partial^2_k\rho_{{\bf
    k}1}^0+2\frac{k}{m}\partial_k\rho_{{\bf k}1}^0,\label{u1}\\
&&\Delta_0\partial_k\rho^0_{{\bf
    k}3}=\frac{k}{m}\rho_{{\bf k}1}^0+\xi_k\partial_k\rho_{{\bf
    k}1}^0,\label{u2}
\end{eqnarray}
Eq.~(\ref{ro2}) becomes
\begin{eqnarray}
\rho^{\omega}_{{\bf
    k}2}&=&\frac{1}{4E^2_k-\omega^2}\Big\{i\omega\delta|\Delta|^{\omega}\frac{\xi_k}{E_k}f_k-i\omega\mu^{\omega}_{\rm 
  eff}\frac{\Delta_0}{E_k}f_k\nonumber\\
&&\mbox{}-\Big[\frac{iE^2_k}{2}({\bf q}\cdot{\bf e_k})({\bf
  p}_s^{\omega}\cdot{\bf e_k})\Big(\frac{k^2\partial^2_{\xi_k}}{m^2}+\frac{\partial_{\xi_k}}{m}\Big)-2\frac{i{\bf k}\cdot{\bf q}}{m}\nonumber\\
&&\mbox{}\times\frac{\bf k}{m}\cdot\frac{{e{\bf
      E}^{\omega}}+\Big({e{\bf
      E}^{\omega}}-i\omega{{\bf p}^{\omega}_s}/{2}\Big)\xi_k\partial_{\xi_k}}{i\omega}\nonumber\\
&&\mbox{}+\frac{i({\bf q}\cdot{\bf e_k})({\bf
  p}_s^{\omega}\cdot{\bf e_k})}{2m}\xi_k-\frac{i{\bf
  q}\cdot{\bf p}^{\omega}_s}{2m}\xi_k\Big]\rho^0_{{\bf k}1}\Big\}.\label{a5}
\end{eqnarray}
Consequently, with $\sum_{\bf k}'\rho^{\omega}_{{\bf k}2}=0$ [Eq.~(\ref{phase})] and $e{\bf E}^{\omega}={i\omega{\bf
    p}^{\omega}_s}/{2}+i{\bf q}\mu^{\omega}_{\rm eff}$, by taking care of the particle-hole symmetry to remove terms
with the odd order of $\xi_k$ in the summation of ${\bf k}$, Eq.~(\ref{phase1}) is obtained.

Particularly, at the low-frequency, i.e.,
$\omega\ll\Delta_0$, the dimensionless factor $s_{\omega}$ in Eq.~(\ref{phase1}) becomes 
\begin{small}
\begin{eqnarray}
s_{\omega}\approx\frac{\sum_{\bf
    k}\big[\frac{1}{2E_k^2}(2-{E_k^2\partial^2_{\xi_k}})\frac{\Delta_0}{E_k}f_k\big]}{\sum_{\bf
    k}\big(\frac{1}{E_k^2}\frac{\Delta_0}{E_k}f_k\big)}=\frac{\sum_{\bf
    k}\big(\frac{\Delta_0}{E^3_k}f_k\big)}{\sum_{\bf
    k}\big(\frac{\Delta_0}{E^3_k}f_k\big)}=1.~~~
\end{eqnarray}
\end{small}

Similarly, $g_{\omega}$ in Eq.~(\ref{phase1}) at low frequency ($\omega\ll\Delta_0$) and low temperature
[$f_k=1-2n_F(E_k)\approx1$] reads:
\begin{eqnarray}
g_{\omega}&=&\frac{\sum_{\bf
    k}\Big[\frac{\Delta_0}{4E_k^2}\partial_{\xi_k}\Big(\frac{\xi_k}{E_k}f_k\Big)\Big]}{\sum_{\bf
    k}\Big(\frac{1}{4E_k^2}\frac{\Delta_0}{E_k}f_k\Big)}\approx\frac{\sum_{\bf
    k}\Big(\frac{\Delta_0}{4E_k^2}\frac{\Delta^2_0}{E_k^3}\Big)}{\sum_{\bf
    k}\Big(\frac{1}{4E_k^2}\frac{\Delta_0}{E_k}\Big)}\nonumber\\
&=&\frac{\int^{\omega_D}_{-\omega_D}\frac{dx}{(1+x^2)^{5/2}}}{\int^{\omega_D}_{-\omega_D}\frac{dx}{(1+x^2)^{3/2}}}\approx\frac{\int^{\infty}_{-\infty}\frac{dx}{(1+x^2)^{5/2}}}{\int^{\infty}_{-\infty}\frac{dx}{(1+x^2)^{3/2}}}=\frac{2}{3}.\label{gw2}
\end{eqnarray}

After sum over ${\bf k}$ in the BCS spherical shell to Eq.~(\ref{a3}), one has
\begin{equation}
-i\omega{\sum_{\bf k}}'\rho^{\omega}_{{\bf k}1}={\sum_{\bf
  k}}'2\xi_k\rho^{\omega}_{{\bf k}2}-{\sum_{\bf k}}'\frac{i{\bf q}\cdot{\bf p}_s^{\omega}}{4m}\rho^{0}_{{\bf k}1}.
\end{equation}
By further using gap equation [Eq.~(\ref{gap})] and the solution of $\rho^{\omega}_{{\bf
    k}2}$ [Eq.~(\ref{a5})], one obtains
\begin{eqnarray}
&&\frac{i\omega\delta|\Delta|^{\omega}}{g}={\sum_{\bf k}}'\frac{2\xi_k}{4E^2_k-\omega^2}\Big\{i\omega\left(\delta|\Delta|^{\omega}{\xi_k}-\mu^{\omega}_{\rm
  eff}{\Delta_0}\right)\frac{f_k}{E_k}\nonumber\\
&&\mbox{}-\Big[\frac{iE^2_k}{2}({\bf q}\cdot{\bf e_k})({\bf
  p}_s^{\omega}\cdot{\bf e_k})\Big(\frac{k^2\partial^2_{\xi_k}}{m^2}+\frac{\partial_{\xi_k}}{m}\Big)-\frac{i{\bf
  q}\cdot{\bf p}^{\omega}_s\xi_k}{3m}\nonumber\\
&&\mbox{}-2\frac{i{\bf k}\cdot{\bf q}}{m}\frac{\bf k}{m}\cdot\frac{\Big({e{\bf
      E}^{\omega}}-i\omega{{\bf p}^{\omega}_s}/{2}\Big)\xi_k\partial_{\xi_k}+{e{\bf
      E}^{\omega}}}{i\omega}\Big]\rho^0_{{\bf k}1}\Big\}\nonumber\\
&&\mbox{}-{\sum_{\bf k}}'\frac{i{\bf q}\cdot{\bf p}_s^{\omega}}{4m}\rho^{0}_{{\bf k}1}.
\end{eqnarray}
Further, by using the particle-hole symmetry to remove terms with the odd
order of $\xi_k$ in the summation of ${\bf k}$, Eq.~(\ref{higgs1}) is obtained. 
$c_{\omega}$ in Eq.~(\ref{higgs1}) is given by $c_{\omega}=z_{\omega}-3\Delta_0/(2g)$ with 
\begin{small}
\begin{eqnarray}
z_{\omega}\!=\!\sum_{\bf k}\frac{[4\xi_k^2(E_k^2\partial^2_{\xi_k}-1+2\xi_k\partial_{\xi_k})+2E_k^2\xi_k\partial_{\xi_k}]\rho^0_{{\bf k}1}}{4E_k^2-\omega^2}.~~~~
\end{eqnarray}
\end{small}

\section{Derivation of Eq.~(\ref{plasma})}
\label{Eff}

We derive the Hartree field [Eq.~(\ref{plasma})] in this part.  Generally,  with the Hartree field (i.e., the vacuum
polarization), the plasma oscillation is involved,  causing the Coulomb screening to the longitudinal
electromagnetic field.  Nevertheless, the transverse field is not affected. 

By first substituting $\rho^{\omega}_{{\bf k}3}$ [Eq.~(\ref{a2})] and then substituting $\rho^{\omega}_{{\bf k}0}$
[Eq.~(\ref{a1})], into
Eq.~(\ref{muH1}), the Hartree field reads:  
\begin{eqnarray}
\mu^{\omega}_H&\!=\!&\frac{2V_q}{\omega}\sum_{\bf
  k}\left[\frac{\left({\bf k}\cdot{\bf q}\right)}{m}\rho^{\omega}_{{\bf
      k}0}\right]\nonumber\\
&\!=\!&-\frac{2V_q}{\omega}\sum_{\bf
  k}\left\{\frac{\left({\bf k}\cdot{\bf q}\right)}{m}\Big[\frac{e{\bf
      E}^{\omega}\cdot{\partial_{\bf k}\rho^0_{{\bf
    k}3}}}{i\omega}\Big]+O(q^2)\right\}\nonumber\\
&\!=\!&\frac{V_qv_F^2{\bf q}\cdot{\bf
    E}^{\omega}}{3i\omega^2}\Big[\sum_{\bf
  k}{\partial_{E_k}f_k}-\sum_{\bf
  k}\frac{\Delta^2_0}{E_k}\partial_{E_k}\Big(\frac{f_k}{E_k}\Big)\Big].~~~~~~\label{AHFD}
\end{eqnarray}
In the superconducting state with $k_BT{\ll}\omega_D$ ($T$ denotes the temperature), one has
$\partial_{E_k}f_k=-2\partial_{E_k}n_F(E_k)\approx0$ when $|\xi_k|>\omega_D$. Therefore, the first summation 
on the right-hand side of Eq.~(\ref{AHFD}) can be restricted inside the spherical shell. Moreover, the second one is also
restricted inside the spherical shell, considering the fact that the gap vanishes outside the spherical shell in the BCS
theory.\cite{BCS,G1} Then, Eq.~(\ref{plasma}) is obtained.

With Eq.~(\ref{plasma}), the linear electric field ${\bf E}^{\omega}$ from Eq.~(\ref{electric}) becomes:
\begin{equation}
{\bf E}^{\omega}={\bf E}^{\omega}_0+\frac{{\bf q}({\bf q}\cdot{{\bf
      E}^{\omega}_0})}{q^2}\frac{{\omega^2_p}/{\omega^2}}{1-{\omega^2_p}/{\omega^2}}. \label{electricf}
\end{equation}
Therefore, it is noted that the longitudinal electric field experiences the Coulomb screening, 
i.e., ${\bf E}^{\omega,\parallel}=\frac{{\bf
    E}^{\omega,\parallel}_0}{1-{\omega^2_p}/{\omega^2}}$ whereas the transverse one does
not (${\bf E}^{\omega,\perp}={\bf E}^{\omega,\perp}_0$), as pointed out above.

\section{Derivation of $n^{2\omega}$, Eqs.~(\ref{sphase1}) and (\ref{shiggs2})}
\label{bb}

We derive Eqs.~(\ref{sphase1}) and (\ref{shiggs2}) in this part. Considering
the long-wave limit, we only keep the lowest two orders of ${\bf q}$ in our
derivation. Then, the second order of the GIKE [Eq.~(\ref{GE})] in the clean limit reads:  
\begin{eqnarray}
&&2i\omega\rho^{2\omega}_{\bf k}+i\left[\xi_k\tau_3+\Delta_0\tau_1,\rho^{2\omega}_{\bf
  k}\right]-i\Big\{\frac{{\bf k}\cdot{\bf q}}{m}\tau_3,\rho^{2\omega}_{\bf
    k}\Big\}\nonumber\\
&&\mbox{}+i\left[\delta|\Delta|^{2\omega}\tau_1+\mu^{2\omega}_{\rm eff}\tau_3,\rho^0_{\bf
    k}\right]-i\Big[\frac{{\bf q}\cdot{\bf p}_s^{2\omega}}{4m}\tau_3,\tau_3\rho^0_{\bf k}\Big]\nonumber\\
&&\mbox{}+\frac{1}{2}\{e{\bf E}^{2\omega}\tau_3+2i{\bf q}\delta|\Delta|^{2\omega}\tau_1-i{\bf
  p}^{2\omega}_s\Delta_0\tau_3\tau_1,\partial_{\bf k}\rho^{0}_{\bf
  k}\}\nonumber\\
&&\mbox{}+\frac{i}{8}[{\bf p}_s^{\omega}{\bf p}_s^{\omega}\Delta_0\tau_1-2{\bf q}{\bf
  p}^{2\omega}_s\Delta_0\tau_3\tau_1,\partial_{\bf k}\partial_{\bf k}\rho^0_{\bf k}]\nonumber\\
&&\mbox{}+\frac{1}{2}\{e{\bf E}^{\omega}\tau_3-i{\bf
  p}^{\omega}_s\Delta_0\tau_3\tau_1,\partial_{\bf k}\rho^{\omega}_{\bf
  k}\}+O(q^2)=0,\label{b0}
\end{eqnarray}
in which we have used the fact that $\delta|\Delta|^{\omega}$
[Eq.~(\ref{higgs2})], $\mu^{\omega}_{\rm eff}$ [Eq.~(\ref{phase1})],
$\rho^{\omega}_{{\bf k}2}$ [Eq.~(\ref{a5})], $\rho^{\omega}_{{\bf k}1}$
[Eq.~(\ref{a3})], $\rho^{\omega}_{{\bf k}3}$ [Eq.~(\ref{a2})] are the
quantities in the first order of $q$. Components of Eq.~(\ref{b0})
can be written as
\begin{small}
\begin{eqnarray}
2i\omega\rho^{2\omega}_{{\bf k}0}&\!=\!&2\frac{i{\bf k}\!\cdot\!{\bf
    q}}{m}\rho^{2\omega}_{{\bf k}3}\!-\!{\partial_{\bf k}}\!\cdot\!(e{\bf
  E}^{\omega}\rho^{\omega}_{{\bf k}3}\!+\!\Delta_0{\bf
  p}_s^{\omega}\rho^{\omega}_{{\bf k}2}\!+\!e{\bf E}^{2\omega}\rho^0_{{\bf
    k}3}\nonumber\\
&&\mbox{}\!+\!2i{\bf q}\rho^0_{{\bf
    k}1}\delta|\Delta|^{2\omega}),\label{b1}\\
2i\omega\rho^{2\omega}_{{\bf k}3}&\!=\!&2\Delta_0\rho^{2\omega}_{{\bf
    k}2}\!+\!2\frac{i{\bf k}\cdot{\bf
    q}}{m}\rho^{2\omega}_{{\bf k}0}\!+\!\frac{i\Delta_0{\bf q}{\bf
  p}_s^{2\omega}\!:\!\partial_{\bf k}\partial_{\bf k}\rho^0_{{\bf k}1}}{2}\nonumber\\
&&\mbox{}\!-\!(e{\bf E}^{\omega}\!\cdot\!{\partial_{\bf k}})\rho^{\omega}_{{\bf k}0},\label{b2}\\
2i\omega\rho^{2\omega}_{{\bf k}1}&\!=\!&\frac{i{\bf
    q}\cdot{\bf p}^{2\omega}_s\rho^{0}_{{\bf k}1}}{2m}\!-\!2\xi_k\rho^{2\omega}_{{\bf
    k}2}\!-\!\frac{i\Delta_0{\bf q}{\bf
  p}_s^{2\omega}\!:\!\partial_{\bf k}\partial_{\bf k}\rho^0_{{\bf k}3}}{2},~~~~~~\label{b3}\\
2i\omega\rho^{2\omega}_{{\bf k}2}&\!=\!&2\xi_k\rho^{2\omega}_{{\bf
    k}1}\!-\!2\Delta_0\rho^{2\omega}_{{\bf k}3}\!+\!({\xi_k}\delta|\Delta|^{2\omega}\!-\!{\Delta_0}\mu^{2\omega}_{\rm
  eff})\frac{f_k}{E_k}\nonumber\\
&&\mbox{}\!-\!\Delta_0({\bf p}_s^{\omega}\cdot{\partial_{\bf k}})\rho^{\omega}_{{\bf k}0}\!-\!\frac{\Delta_0}{4}{\bf
  p}_s^{\omega}{\bf p}_s^{\omega}\!:\!\partial_{\bf k}\partial_{\bf k}\rho^0_{{\bf k}3}.\label{b4}
\end{eqnarray}
\end{small}

Then, by first substituting Eq.~(\ref{a1}) and then substituting Eqs.~(\ref{b2})
and (\ref{b3}) into Eq.~(\ref{b4}), $\rho^{2\omega}_{{\bf k}2}$ can be obtained:
\begin{eqnarray}
&&\rho^{2\omega}_{{\bf
    k}2}=\frac{1}{4(E^2_k-\omega^2)}\Big\{4i\omega(\mu^{2\omega}_{\rm eff}\rho^0_{{\bf
    k}1}-\delta|\Delta|^{2\omega}\rho^0_{{\bf k}3})-2i\omega\Delta_0\nonumber\\
&&\mbox{}\times\frac{\big({e{\bf
        E}^{\omega}}-{i\omega}{\bf p}^{\omega}_s\big)\cdot{\partial_{\bf k}}}{i\omega}\frac{e{\bf
        E}^{\omega}\cdot{\partial_{\bf k}}}{i\omega}\rho^0_{{\bf k}3}+\frac{i{\bf
    q}\cdot{\bf p}^{2\omega}_s\xi_k\rho^0_{{\bf k}1}}{m}\nonumber\\
&&\mbox{}-\Delta_0\Big[i({\bf q}\cdot{\bf e_k})({\bf p}^{2\omega}_s\cdot{\bf
  e_k})(\Delta_0\partial_k^2\rho^0_{{\bf k}1}+\xi_k\partial_k^2\rho^0_{{\bf k}3})+\frac{\partial^2_k\rho^0_{{\bf
    k}3}}{2}\nonumber\\
&&\mbox{}\times{i}\omega({\bf
  p}_s^{\omega}\cdot{\bf e}_{\bf k})^2-2i\frac{{\bf k}\cdot{\bf q}}{m}\frac{e{\bf
    E}^{2\omega}\cdot\partial_{\bf k}\rho^0_{{\bf k}3}}{i\omega}\Big]\Big\}+O(q^2),\label{bb1}
\end{eqnarray}
in which Eq.~(\ref{b1}) is used for $\rho^{2\omega}_{{\bf
    k}0}$. With the help of Eqs.~(\ref{u1}) and (\ref{u2}), by considering  $e{\bf
  E}^{2\omega}=i\omega{\bf p}^{2\omega}_s+2i{\bf q}{\mu^{2\omega}_{\rm eff}}$, Eq.~(\ref{bb1}) becomes
\begin{eqnarray}
&&\rho^{2\omega}_{{\bf
    k}2}=\frac{1}{4(E^2_k-\omega^2)}\Big\{4i\omega(\mu^{2\omega}_{\rm eff}\rho^0_{{\bf
    k}1}-\delta|\Delta|^{2\omega}\rho^0_{{\bf k}3})\nonumber\\
&&\mbox{}-2i\omega\Delta_0\Big[\big(\frac{e{\bf
        E}^{\omega}}{i\omega}-{\bf p}^{\omega}_s\big)\cdot{{\bf e_k}}\Big]^2\Big(\frac{k^2\partial^2_{\xi_k}}{m^2}+\frac{\partial_{\xi_k}}{m}\Big)\rho^0_{{\bf
    k}3}\nonumber\\
&&\mbox{}-2i\omega\Delta_0\frac{({e{\bf
        E}^{\omega}}-i\omega{\bf p}^{\omega}_s)\cdot{\bf
    e_{\theta_k}}}{i\omega}\frac{e{\bf
        E}^{\omega}\cdot{\bf
      e_{\theta_k}}}{i\omega}\frac{\partial_{\xi_k}\rho^0_{{\bf k}3}}{m}\nonumber\\
&&\mbox{}-4\Big(\frac{{\bf k}\cdot{\bf
      q}}{m}\Big)^2\frac{\mu^{2\omega}\partial_{\xi_k}(\xi_k\rho^0_{{\bf
k}1})}{i\omega}+2i\frac{{\bf k}\cdot{\bf
      q}}{m}\frac{{\bf k}\cdot{\bf
      p}_s^{2\omega}}{m}\rho^0_{{\bf k}1}\nonumber\\
&&\mbox{}-iE^2_k({\bf
    q}\cdot{\bf e_k})({\bf p}^{2\omega}_s\cdot{\bf
    e_k})\Big(\frac{k^2\partial^2_{\xi_k}}{m^2}+\frac{\partial_{\xi_k}}{m}\Big)\rho^0_{{\bf
    k}1}\nonumber\\
&&\mbox{}+i\Big[\frac{{\bf
    q}\cdot{\bf p}^{2\omega}_s}{m}-\frac{({\bf
    q}\cdot{\bf e_k})({\bf p}^{2\omega}_s\cdot{\bf
    e_k})}{m}\Big]\xi_k\rho^0_{{\bf k}1}\Big\}. \label{d4}
\end{eqnarray} 
Then, with  $\sum_{\bf k}'\rho^{2\omega}_{{\bf k}2}=0$ [Eq.~(\ref{phase})], via taking care of the particle-hole symmetry
to remove terms with the odd order of $\xi_k$ in the summation of ${\bf k}$, Eq.~(\ref{sphase1}) is obtained.
Particularly, in Eq.~(\ref{sphase1}), the dimensionless factor $l_{\omega}$ [Eq.~(\ref{lw})] at low frequency and low
temperature reads:
\begin{eqnarray}
l_{\omega}&\approx&\frac{\sum_{\bf k}\Big[\frac{\Delta_0}{E_k^2}(2\xi_k\partial^2_{\xi_k}+\partial_{\xi_k})\Big(\frac{\xi_k}{E_k}\Big)\Big]}{3\sum_{\bf
  k}\Big(\frac{1}{E_k^2}\frac{\Delta_0}{E_k}\Big)}=\frac{\sum_{\bf
  k}\Big(\frac{\Delta^3_0}{E^5_k}-6\frac{\xi^2_k\Delta^3_0}{E^6_k}\Big)}{3\sum_{\bf
k}\frac{\Delta_0}{E^3_k}}\nonumber\\
&\approx&\frac{\int^{\omega_D}_{-\omega_D}dx[\frac{1}{(1+x^2)^{5/2}}-\frac{6x^2}{(1+x^2)^{7/2}}]}{3\int^{\omega_D}_{-\omega_D}\frac{dx}{(1+x^2)^{3/2}}}=-\frac{2}{45}. 
\end{eqnarray}

Following the derivation of the linear $\mu_H^{\omega}$ above,  by substituting $\rho^{2\omega}_{{\bf k}3}$ [Eq.~(\ref{b2})] into the second-order Hartree
and Fock fields [Eqs.~(\ref{muH2}) and (\ref{muF2})], one has  
\begin{eqnarray}
&&\mu^{2\omega}_H+\mu^{2\omega}_F\!=\!{(V_{2q}+\frac{g}{2})n^{2\omega}}\!=\!\frac{2V_{2q}+g}{\omega}\sum_{\bf
  k}\Big[\frac{\left({\bf k}\cdot{\bf q}\right)}{m}\rho^{2\omega}_{{\bf
      k}0}\Big]\nonumber\\
&&\mbox{}\!=\!-\frac{2V_{2q}+g}{2i\omega}\sum_{\bf
  k}\Big\{\frac{({\bf k}\cdot{\bf q})}{m}\Big[\frac{e{\bf
      E}^{2\omega}\cdot{\partial_{\bf k}\rho^0_{{\bf
    k}3}}}{\omega}\Big]+O(q^2)\Big\}\nonumber\\
&&\mbox{}\!=\!\frac{2V_{2q}+g}{2im\omega^2}\frac{({\bf q}\cdot{\bf
    E}^{2\omega})}{2}(\rho_Q+\rho_s),\label{d5}
\end{eqnarray}
with $\rho_Q=\frac{k_F^2}{3m}\sum_{\bf
  k}{\partial_{E_k}f_k}$.
Further substituting the second-order electric field $e{\bf E}^{2\omega}=2i{\bf
  q}(\mu^{2\omega}_H+\mu_F^{2\omega})$ [Eq.~(\ref{electric})] into Eq.~(\ref{d5}), one obtains
\begin{equation}
\mu^{2\omega}_H+\mu_F^{2\omega}=(\mu^{2\omega}_H+\mu_F^{2\omega})\frac{q^2(2V_{2q}+g)(\rho_Q+\rho_s)}{2m\omega^2}.
\end{equation}
Therefore, one immediately finds the vanishing 
$\mu^{2\omega}_H$, $\mu^{2\omega}_F$, $e{\bf E}^{2\omega}$ and $n^{2\omega}$. 

After the summation of ${\bf k}$ in the BCS spherical shell to Eq.~(\ref{b3}), one comes to
\begin{equation}
-2i\omega{\sum_{\bf k}}'\rho^{2\omega}_{{\bf k}1}={\sum_{\bf
  k}}'2\xi_k\rho^{2\omega}_{{\bf k}2}-{\sum_{\bf k}}'\frac{i{\bf q}\cdot{\bf p}_s^{2\omega}}{2m}\rho^{0}_{{\bf k}1}.\label{c8}
\end{equation}
Considering the long-wave limit ($q=0$), the second term on the right-hand side of Eq.~(\ref{c8}) vanishes. Then,
substituting the solution of  $\rho^{2\omega}_{{\bf k}2}$ [Eq.~(\ref{d4})] which is simplified at $q=0$ into Eq.~(\ref{c8}), one obtains
\begin{eqnarray}
&&-{\sum_{\bf k}}'\rho^{2\omega}_{{\bf k}1}={\sum_{\bf
  k}}'\frac{\xi_k}{E^2_k-\omega^2}\Big\{\mu^{2\omega}_{\rm eff}\rho^0_{{\bf
    k}1}-\delta|\Delta|^{2\omega}\rho^0_{{\bf k}3}\nonumber\\
&&\mbox{}-\Big[\Big(\frac{e{\bf
        E}^{\omega}}{i\omega}-{\bf p}^{\omega}_s\Big)\cdot{{\bf e_k}}\Big]^2\frac{\Delta_0}{2}\Big(\frac{k^2\partial^2_{\xi_k}}{m^2}+\frac{\partial_{\xi_k}}{m}\Big)\rho^0_{{\bf
    k}3}\nonumber\\
&&\mbox{}-\frac{\Delta_0}{2}\Big[\Big(\frac{e{\bf
        E}^{\omega}}{i\omega}-{\bf p}^{\omega}_s\Big)\cdot{\bf
    e_{\theta_k}}\Big]\Big(\frac{e{\bf
        E}^{\omega}}{i\omega}\cdot{\bf
      e_{\theta_k}}\Big)\frac{\partial_{\xi_k}\rho^0_{{\bf k}3}}{m}\Big\}.~~~~~~~
\end{eqnarray}
By further using the gap equation [Eq.~(\ref{gap})] and taking care
of the particle-hole symmetry to remove terms with the odd order of $\xi_k$ in the summation of ${\bf k}$, one directly
obtains Eq.~(\ref{shiggs2}).

\section{Derivation of Eq.~(\ref{TD5})}
\label{ee}

For excitation with $\theta_{L}=z^{L}_{\omega_{L}}\cos(2\omega_{L}t)$ and
$\theta_{R}=z^{R}_{\omega_{R}}\sin(2\omega_{R}t)$ in Fig.~\ref{figyw2}, the dc-current component in the induced
Josephson current $J=J_c\sin(\theta_L-\theta_R)$ can be obtained through a time 
average:
\begin{widetext}
  \begin{eqnarray}
    J^{\rm dc}&=&\frac{1}{T}\int^T_0J=\frac{1}{T}\int^T_0J_c[\sin(z^L_{\omega_{L}}\cos2\omega_Lt)\cos(z^R_{\omega_{R}}\sin2\omega_Rt)-\cos(z^L_{\omega_{L}}\cos2\omega_Lt)\sin(z^R_{\omega_{R}}\sin2\omega_Rt)]\nonumber\\
    &=&\frac{1}{T}\int^T_0J_c\Big\{\big\{-2\sum^{\infty}_{n=1}(-1)^nj_{2n-1}(z^L_{\omega_{L}})\cos[(2n-1)2\omega_Lt]\big\}\big\{j_0(z^R_{\omega_{R}})+2\sum^{\infty}_{m=1}j_{2m}(z^R_{\omega_{R}})\cos[(2m)2\omega_Rt]\big\}\nonumber\\
    &&\mbox{}-\big\{j_0(z^L_{\omega_{L}})+2\sum^{\infty}_{m=1}(-1)^mj_{2m}(z^L_{\omega_{L}})\cos[(2m)2\omega_Lt]\big\}\big\{2\sum^{\infty}_{n=1}j_{2n-1}(z^R_{\omega_{R}})\sin[(2n-1)2\omega_Rt]\big\}\Big\}\nonumber\\
    &=&2J_c\sum^{\infty}_{n=1}\sum^{\infty}_{m=1}(-1)^{n+1}\Big[j_{2n-1}(z^L_{\omega_{L}})j_{2m}(z^R_{\omega_{R}})\delta_{(2n-1)\omega_L,(2m)\omega_R}\Big].\label{TD1}
  \end{eqnarray}
\end{widetext}
Particularly, for the weak phase excitation (small $z^l$ and $z^R$), only the lowest two orders of the Bessel function
are important, and then, Eq.~(\ref{TD5}) is obtained.

\end{appendix}


\begin{thebibliography}{0}
\bibitem{AKs} In fact, by looking into Ref.~\onlinecite{AK}, the charge conservation [Eq.~(2.12)] is not justified in
  the superconducting state with the spontaneous breaking of the $U(1)$ symmetry but discussed before the mean-field
  theory where the $U(1)$ symmetry still holds. In Ref.~\onlinecite{AK},
  to derive the NG mode [Eq.~(3.35)] after the mean-field theory, an additional condition of the charge conservation is
  required besides the Gorkov's equation.
\end{thebibliography}
\end{document}